\documentclass[aps,twocolumn,showpacs,prx,amsmath,amssymb,floatfix,superscriptaddress]{revtex4-1}
\usepackage{amsmath,amssymb,graphicx,color}
\usepackage{bm}
\usepackage{bbm}
\usepackage[caption=false]{subfig}
\usepackage[usenames,dvipsnames]{xcolor}
\usepackage[colorlinks=true,citecolor=Cerulean,linkcolor=RubineRed,urlcolor=Cerulean]{hyperref}
\usepackage[normalem]{ulem}
\usepackage{soul,xcolor}
\usepackage{multirow}

\newcommand{\avg}[1]{\ensuremath{\langle #1 \rangle}}

\usepackage{fixltx2e,amsmath}
\MakeRobust{\eqref}

\usepackage{mathptmx} 

\DeclareMathAlphabet{\mathcal}{OMS}{cmsy}{m}{n}
\DeclareSymbolFont{largesymbols}{OMX}{cmex}{m}{n}

\begin{document}
\setstcolor{blue}
\title{Unified geometric formalism for dissipation and its fluctuations in finite-time microscopic heat engines}

\author{Gentaro Watanabe}
\affiliation{School of Physics and Zhejiang Institute of Modern Physics, Zhejiang University, Hangzhou, Zhejiang 310027, China}
\affiliation{Zhejiang Province Key Laboratory of Quantum Technology and Device, Zhejiang University, Hangzhou, Zhejiang 310027, China}
\author{Guo-Hua Xu}
\affiliation{Universal Biology Institute, The University of Tokyo,
7-3-1 Hongo, Bunkyo-ku, Tokyo 113-0033, Japan}
\author{Yuki Minami}
\affiliation{Faculty of Engineering, Gifu University, Yanagido, Gifu 501-1193, Japan}

\date{\today}

\begin{abstract}
Microscopic heat engines operate in regimes where thermodynamic quantities fluctuate strongly, making stochastic effects an essential aspect of their performance. However, existing geometric formulations of finite-time thermodynamics primarily characterize average dissipation and do not systematically capture its fluctuations. Here, we develop a unified geometric framework that consistently describes both the mean dissipated availability and its fluctuations. In the linear-response regime, we show that these quantities are governed by metric tensors constructed from equilibrium correlation functions, providing a common geometric structure for dissipation and its fluctuations. This framework yields geometric bounds on both the mean and variance of the dissipated availability, and thereby on the efficiency and its fluctuations. The formalism applies broadly to stochastic systems, including Markov jump processes and overdamped and underdamped Brownian dynamics, establishing a unified geometric description across microscopic heat engines.
\end{abstract}

\maketitle

\section{Introduction}

Recent technological advances have enabled the realization and precise control of thermal machines at microscopic scales~\cite{Bustamante05,Ciliberto17}. Examples include heat engines using a single or a small number of microparticles or ions~\cite{Martinez17,Blickle12,Quinto-Su14,Martinez16,Rossnagel16,Argun17,Maslennikov19,vonLindenfels19,Hou25}, mesoscopic electronic devices~\cite{Steeneken11,Pekola15}, and nanomechanical devices~\cite{Hugel02,Klaers17}, in which thermodynamic cycles can be implemented and monitored at the level of individual degrees of freedom. Since the machines operate far from the macroscopic limit, thermodynamic quantities such as work, heat, and efficiency exhibit pronounced fluctuations, and performance is typically constrained by finite-time operation. In particular, fluctuations of efficiency play a crucial role in assessing the performance and reliability of microscopic heat engines. These developments have renewed interest in finite-time thermodynamics of small systems~\cite{Seifert12}, not only as a tool for optimizing the average performance of small thermal machines, but also for understanding the role of fluctuations in realistic cyclic processes.

Motivated by these experimental realizations, stochastic thermodynamics has provided a powerful framework to describe thermodynamic processes in small systems~\cite{Sekimoto98,Sekimotobook10,Pelitibook21,Saitobook22,Shiraishibook23,Seifertbook25,Jarzynski11,Seifert12,VandenBroeck14,Nicolis17,Guery-Odelin23}. (Applications of stochastic thermodynamics also include systems involving biological and active matter reservoirs~\cite{Krishnamurthy16,Pietzonka19,Fodor21,Toyabe15}.) This framework enables a consistent definition of thermodynamic quantities along individual stochastic trajectories and has clarified the role of fluctuations in nonequilibrium processes. Within this framework, fluctuation theorems~\cite{Jarzynski97,Crooks99,Jarzynski11,Seifert12} and related universal relations, such as thermodynamic uncertainty relations (TURs)~\cite{Barato15,Gingrich16,Seifert19,Hasegawa19,Horowitz20}, impose general constraints on nonequilibrium fluctuations. TURs have also been extended to time-dependent processes~\cite{Liu20,Koyuk20,Tan20}, including periodically driven systems~\cite{Barato18,Koyuk19a,Koyuk19b,Miller21,Wang24}, underscoring the fundamental role of stochasticity in small-scale thermodynamic processes. In the context of microscopic heat engines, stochastic thermodynamics has been widely used to analyze both average performance and the statistics of work and heat~\cite{Sekimoto00,VandenBroeck05,Schmiedl08,Esposito10,Dechant15,Brandner15,Dechant17,Strasberg21,  Hoppenau13,Holubec14,Rana14,Cerino15,Holubec17,Holubec18,Dechant19,Saryal21,Mohanta21,Holubec21,   Verley14a,Verley14b,Proesmans15a,Proesmans15b,Park16,Saha18,Manikandan19,Vroylandt20,Sinitsyn11,Pal17,   Proesmans17,Pietzonka18,Barato18,Koyuk19a,Koyuk19b,Timpanaro19,Kamijima21,   Plata19,Plata20a,Plata20b, Xu21,Xu22,Ito25}, including experimental investigations of work and dissipation fluctuations~\cite{Jop08,Martinez15}. However, while stochastic thermodynamics offers general principles, the systematic characterization of dissipation and its fluctuations in driven cyclic processes remains challenging, particularly when one seeks descriptions that are both transparent and broadly applicable across different dynamical regimes.

To address these challenges, geometric formulations of finite-time thermodynamics have emerged as a promising approach~\cite{Weinhold75,Weinhold76,Ruppeiner79,Ruppeiner95,Andresen11}. In these formulations, the dissipation generated during finite-time thermodynamic processes is related to geometric quantities, such as the length of a path in control-parameter space, which allows one to derive relations that hold independently of the specific details of the driving protocol~\cite{Salamon83,Andresen84,Nulton85,Salamon84,Gilmore84,Schlogl85,Brody94,Crooks07,Zulkowski12,Vu21,Eglinton22,Frim22prl,Sawchuk26}. In the linear-response regime, Sivak and Crooks demonstrated that the average excess work can be expressed in terms of a thermodynamic metric defined by equilibrium correlation functions~\cite{Sivak12}, and later such geometric ideas were successfully applied to closed thermodynamic cycles~\cite{Brandner20,Abiuso20}. More recently, Frim and DeWeese, as well as Li and Izumida, developed related geometric descriptions of finite-time heat engines for a specific system~\cite{Frim22,Li25}. Furthermore, for systems characterized by a single correlation time, such as an overdamped Brownian particle in a harmonic potential, it has been shown that not only the average dissipation accumulated over a cycle but also its variance can be described within a geometric framework~\cite{Watanabe22} (work fluctuations and related issues have also been discussed using geometric approaches; see, e.g., Refs.~\cite{Miller19,Miller20}). Nevertheless, despite these advances, existing geometric formulations remain largely restricted to average quantities and/or specific dynamical settings, such as systems with a single relaxation timescale. In microscopic heat engines, however, thermodynamic quantities fluctuate strongly, and fluctuations of dissipation play a crucial role in determining performance. This highlights the need for a unified geometric framework that consistently accounts for both the mean and variance of dissipation, and that applies across different dynamical regimes---from overdamped to underdamped systems with multiple relaxation timescales.

In this work, we address this gap by developing a unified geometric framework that simultaneously characterizes the mean and fluctuations of dissipation in finite-time cyclic heat engines. Our formulation applies consistently to both overdamped and underdamped dynamics and naturally accommodates systems with multiple relaxation timescales. In particular, it generalizes our previous formulation for systems with a single correlation time~\cite{Watanabe22} to systems characterized by multiple relaxation timescales, thereby providing a unified framework for describing both the mean and fluctuations of dissipation in finite-time cyclic heat engines. Here, ``finite time'' in the present work refers to finite but sufficiently slow driving, such that a linear-response description remains valid. Within a linear-response, time-local description, we show that both the average dissipation and its variance admit geometric representations governed by metric tensors constructed from common equilibrium correlation functions. Importantly, our framework enables an explicit derivation of the dynamical content of these metrics directly from the underlying correlation functions.

The structure of this paper is as follows. In Sec.~\ref{sec:formalism}, we present our theoretical formalism, which allows us to determine the metrics for the average and fluctuation of dissipation from equilibrium two-time correlation functions. We then apply this general formalism to various systems in the forthcoming sections (Secs.~\ref{sec:markovjump}--\ref{sec:underdampedbp}). In Sec.~\ref{sec:markovjump}, we consider Markov jump systems with discrete states and, in particular, a classical two-level system as a typical example. In Sec.~\ref{sec:overdampedbp}, we discuss an overdamped Brownian particle in a one-dimensional (1D) power-law potential. In particular, the systems discussed in Secs.~\ref{sec:overdampedbp} and \ref{sec:underdampedbp} possess multiple relaxation timescales, requiring the extended formalism developed in the present work. In Sec.~\ref{sec:underdampedbp}, we discuss an underdamped Brownian particle in a harmonic potential as an illustrative example of an underdamped system. Finally, we conclude the paper in Sec.~\ref{sec:conclusion}.

\section{General formalism}\label{sec:formalism}

\subsection{Setup}\label{sec:setup}

We consider a classical microscopic heat engine whose working substance is continuously in contact with a thermal environment with a controllable temperature $T$. The working substance is described by the Hamiltonian $H_{\lambda_w}$, which includes an externally controlled mechanical parameter $\lambda_w$; for instance, in the archetypical setup of a gas confined in a cylinder with a movable piston, $\lambda_w$ corresponds to the position of the piston and determines the volume $V$ of the gas. Thus, the system has two controllable parameters, $\lambda_w$ and $\lambda_u \equiv T$, and the parameter space is spanned by the two-dimensional vector $\lambda_\mu \equiv (\lambda_w, \lambda_u)$. These parameters can also be regarded as generalized displacements.

The natural thermodynamic variables for this system are $(V, T)$, and the corresponding thermodynamic potential is the Helmholtz free energy $F$. Since the total differential of $F$ is given by
\begin{align}
  dF = -S dT - P dV\,,
\end{align}
where $S$ is the entropy and $P$ is the pressure. For the choice of generalized displacements $\lambda_\mu = (V, T)$, the conjugate generalized forces are defined as $X_\mu \equiv (X_w, X_u) = (P,S)$. At the microscopic level, these generalized forces can be expressed by random variables as
\begin{align}
  X_w = -\frac{\partial H_{\lambda_w}}{\partial \lambda_w}\,,\\
  X_u = -k_{\mathrm{B}} \ln{\mathcal{P}}\,,
\end{align}
where $\mathcal{P}=\mathcal{P}(\Gamma, t)$ denotes the probability distribution function over microstates $\Gamma$ and $k_{\mathrm{B}}$ is the Boltzmann constant. In discrete-state Markov jump systems, the distribution $\mathcal{P}(\Gamma, t)$ reduces to state probabilities $P_i(t)$ of state $\Gamma=i$. For continuous systems such as Brownian particles, $\mathcal{P}(\Gamma, t)$ becomes the phase-space distribution $\rho(x, p, t)$ for a phase-space point $\Gamma=(x, p)$, where $x$ and $p$ are the position and the momentum of the particle, respectively.

In the following, the ensemble average with respect to the probability distribution function $\mathcal{P}$ is denoted by
\begin{equation}
  \avg{\cdots} \equiv \int d\Gamma\, \cdots \mathcal{P}(\Gamma)\,.
\end{equation}
In particular, the average in the equilibrium state is denoted by $\avg{\cdots}_{\mathrm{eq}} \equiv \int d\Gamma\, \cdots \mathcal{P}_{\mathrm{eq}}(\Gamma)$, where $\mathcal{P}_{\mathrm{eq}}$ is the equilibrium distribution function. Fluctuations of a random variable $X$ are defined as $\Delta X \equiv X - \avg{X}$.

The control parameters $\lambda_\mu$ are varied cyclically along a closed path in the parameter space with cycle period $\tau$. After sufficiently many cycles, the system reaches a periodic state in which the probability distribution satisfies
\begin{equation}
  \mathcal{P}(\Gamma, t+\tau) = \mathcal{P}(\Gamma, t)
\end{equation}
for any $t \ge 0$. The time origin $t=0$ is chosen within this periodic state.

\subsection{Dissipated availability}

Energy dissipation due to finite-speed driving is quantified by the dissipated availability $A$ defined as~\cite{Salamon83}
\begin{equation}
  A \equiv U - W\,,\label{eq:dissipavail}
\end{equation}
where $W$ is the work output by the engine and $U$ is the effective thermal energy input from the environment (not to be confused with the internal energy). Here, $A$, $W$, and $U$ are random variables given by
\begin{align}
  W \equiv& \oint_\mathcal{C} P\, dV = \int_0^\tau dt \left(- \frac{\partial H_{\lambda_w}}{\partial \lambda_w} \right) \dot{\lambda}_w = \int_0^\tau dt\, X_w \dot{\lambda}_w\,,\label{eq:w}\\
  U \equiv& \oint_\mathcal{C} T\, dS = \int_0^\tau dt\, T \frac{d}{dt}(- k_{\mathrm{B}} \ln{\mathcal{P}}) = \int_0^\tau dt\, \lambda_u \dot{X}_u\,,\label{eq:u}
\end{align}
where $\mathcal{C}$ is the closed path in the parameter space for the engine cycle and the dot denotes the time derivative. In the quasistatic limit, $U$ coincides with the heat input and $A$ vanishes. Thus, $A$ can be interpreted as the part of the effective thermal energy input that cannot be converted into useful work because of finite-time irreversibility.

\subsection{Equilibrium correlations and time-local approximation}

In general, the equilibrium two-time correlation functions $\avg{\Delta X_\mu(t)\, \Delta X_\nu(t')}_{\mathrm{eq}}$, with $\mu$, $\nu \in \{w,\, u\}$, reflect the relaxation properties of the underlying microscopic dynamics and may involve multiple characteristic timescales. Such a situation naturally arises in generic Markov processes, including both discrete-state jump systems and continuous Brownian dynamics.

A particularly simple case is when the correlation function is governed by a single exponential decay. This situation was analyzed in Ref.~\cite{Watanabe22} for an overdamped Brownian particle trapped in a 1D harmonic potential,
\begin{align}
  V_{\lambda_w}(x) = \frac{\lambda_w}{2} x^2\,,\label{eq:vho}
\end{align}
where $x$ is the position of the Brownian particle. In this case, the equilibrium two-time correlation function of the fluctuations of generalized forces $X_\mu$ takes the form
\begin{align}
  \avg{\Delta X_\mu(t)\, \Delta X_\nu(t')}_{\rm eq} = \avg{\Delta X_\mu(t)\, \Delta X_\nu(t)}_{\mathrm{eq}}\, e^{-|t-t'|/\tau_{\mu\nu}}\,,\label{eq:corrsingle}
\end{align}
with the correlation time
\begin{align}
  \tau_{\mu\nu} = \frac{\gamma}{2 \lambda_w}\,,
\end{align}
where $\gamma$ is the friction coefficient. In the slow-driving regime and on coarse-grained timescales, this correlation function can be approximated by a time-local expression,
\begin{align}
  \avg{\Delta X_\mu(t)\, \Delta X_\nu(t')}_{\mathrm{eq}} = 2 \tau_{\mu\nu}(t)\, \avg{\Delta X_\mu(t)\, \Delta X_\nu(t)}_{\mathrm{eq}}\, \delta(t-t')\,.\label{eq:corrod}
\end{align}
This approximation is valid provided that the characteristic driving timescale is sufficiently long compared with the correlation times $\tau_{\mu\nu}$, so that the detailed time dependence of the correlation function becomes irrelevant.

Motivated by this observation, we now consider the general case in which the equilibrium two-time correlation function consists of multiple exponential modes with different relaxation timescales $\tau_{\mu\nu}^{(i)}$:
\begin{align}
  \avg{\Delta X_\mu(t) \Delta X_\nu(t')}_{\mathrm{eq}} = \avg{\Delta X_\mu(t) \Delta X_\nu(t)}_{\mathrm{eq}}\, \sum_{i} C_{\mu\nu}^{(i)}\, e^{-|t-t'|/\tau_{\mu\nu}^{(i)}}\,,\label{eq:corrmulti}
\end{align}
with the normalization condition
\begin{align}
  \sum_i C_{\mu\nu}^{(i)} = 1\,.
\end{align}
Here, $\tau_{\mu\nu}^{(i)}$ are in general complex, corresponding to damped oscillatory modes of the dynamics, as well as purely real in the case of simple overdamped relaxation.

In the slow-driving regime and on coarse-grained timescales, the correlation function can again be approximated by a time-local form
\begin{align}
  \avg{\Delta X_\mu(t)\, \Delta X_\nu(t')}_{\mathrm{eq}} = 2 \overline{\tau}_{\mu\nu}(t)\, \avg{\Delta X_\mu(t)\, \Delta X_\nu(t)}_{\mathrm{eq}}\, \delta(t-t')\,,\label{eq:corrgen}
\end{align}
but now with an effective correlation time given by the weighted sum:
\begin{align}
  \overline{\tau}_{\mu\nu} \equiv \sum_i C_{\mu\nu}^{(i)}\, \tau_{\mu\nu}^{(i)}\,.\label{eq:taubar}
\end{align}
This approximation provides a minimal prescription to incorporate microscopic timescales into thermodynamics through the effective correlation time $\overline{\tau}_{\mu\nu}$, independently of the detailed structure of the underlying dynamics.

The above approximation can be justified within linear-response theory. The metric tensor $g^{(1)}_{\mu\nu}$ governing the mean value $\avg{A}$ of the dissipated availability for a given cycle,
\begin{equation}
  \avg{A} = \int_0^\tau dt\, g^{(1)}_{\mu\nu}(t) \dot{\lambda}_\mu(t) \dot{\lambda}_\nu(t)\,,\label{eq:avra}
\end{equation}
is given by~\cite{Brandner20,Watanabe22,Frim22}
\begin{align}
  g^{(1)}_{\mu\nu}(t) = \beta(t) \int_0^\infty dt'\, \avg{\Delta X_\mu(t+t')\, \Delta X_\nu(t)}_{\mathrm{eq}}\,,\label{eq:g1lrt}
\end{align}
where $\beta(t) \equiv 1/k_{\mathrm{B}}T(t)$ is the instantaneous inverse temperature of the thermal environment.
Substituting the exact expression (\ref{eq:corrmulti}) for the correlation function into Eq.~(\ref{eq:g1lrt}), we obtain
\begin{align}
  g^{(1)}_{\mu\nu}(t)
  &= \beta(t)\, \avg{\Delta X_\mu(t)\, \Delta X_\nu(t)}_{\rm eq}\, \sum_i C_{\mu\nu}^{(i)} \tau^{(i)}_{\mu\nu}\nonumber\\
  &= \beta(t)\, \overline{\tau}_{\mu\nu}(t)\, \sigma_{\mu\nu}(t)\,,\label{eq:g1munu}
\end{align}
where
\begin{equation}
  \sigma_{\mu\nu}(t) \equiv \avg{\Delta X_\mu(t)\, \Delta X_\nu(t)}_{\mathrm{eq}}
\end{equation}
is the covariance tensor of $X_\mu$ and $X_\nu$.

On the other hand, from the approximate expression (\ref{eq:corrgen}) of the correlation function, we obtain
\begin{align}
  g^{(1)}_{\mu\nu}(t) &\simeq \beta(t)\, \int_0^\infty dt'\, 2 \overline{\tau}_{\mu\nu}(t) \avg{\Delta X_\mu(t)\, \Delta X_\nu(t)}_{\rm eq}\, \delta(t') \nonumber\\
  &= \beta(t)\, \overline{\tau}_{\mu\nu}(t)\, \sigma_{\mu\nu}(t)\,,\label{eq:g1munu_final}
\end{align}
where we have used $\int_0^a dt\, \delta(t) = 1/2$ for an arbitrary positive real number $a$. This confirms the consistency of the approximation consisting of Eqs.~(\ref{eq:corrgen}) and (\ref{eq:taubar}).

\subsection{Unified geometric structure of mean and variance of dissipation}

Having established the metric structure governing the mean dissipation, we now show that its fluctuations obey a closely related geometric description. Once the metric $g^{(1)}_{\mu\nu}$ for the mean value $\avg{A}$ of the dissipated availability has been obtained, we now turn to its fluctuations. Under the linear-response and time-local approximations, the variance $\avg{\Delta A^2}$ of $A$ can also be expressed in a metric form similar to Eq.~(\ref{eq:avra}) for $\avg{A}$ (see Appendix~\ref{app:derivation} for details)~\cite{Watanabe22}.
The variance of $A=U-W$ can be decomposed as $\avg{\Delta A^2} = \avg{\Delta U^2} + \avg{\Delta W^2} - 2 \avg{\Delta U\, \Delta W}$. Each term can be expressed as a double time integral involving the two-time correlation functions of the corresponding generalized forces. Applying the time-local approximation introduced above, the double integrals reduce to a single integral over the driving protocol. As shown in Ref.~\cite{Watanabe22}, all three contributions share the same quadratic structure in the driving velocities $\dot{\lambda}_\mu$.

Consequently, the variance of $A$ can be written in the metric form
\begin{align}
  \avg{\Delta A^2} = \int_0^\tau dt\, g^{(2)}_{\mu\nu}(t)\, \dot{\lambda}_\mu(t)\, \dot{\lambda}_\nu(t)\,,\label{eq:vara}
\end{align}
with
\begin{align}
  g^{(2)}_{\mu\nu}(t) \equiv 2\, \overline{\tau}_{\mu\nu}(t)\, \avg{\Delta X_\mu(t) \Delta X_\nu(t)}_{\mathrm{eq}} = 2\, \overline{\tau}_{\mu\nu}(t)\, \sigma_{\mu\nu}(t)\,.\label{eq:g2munu}
\end{align}
A detailed derivation of Eqs.~(\ref{eq:vara}) and (\ref{eq:g2munu}) is provided in Appendix~\ref{app:derivation}. Notably, the metric tensor $g^{(1)}_{\mu\nu}$ appearing in the average of the dissipated availability and $g^{(2)}_{\mu\nu}$ governing its variance are constructed from identical equilibrium correlation functions.

Finally, by comparing between Eqs.~(\ref{eq:g1munu}) and (\ref{eq:g2munu}), we note that the two metrics satisfy the following relation analogous to the fluctuation-dissipation relation~\cite{Watanabe22}:
\begin{equation}
  g^{(2)}_{\mu\nu}(t) = 2k_{\mathrm{B}}T(t) g^{(1)}_{\mu\nu}(t)\,.\label{eq:g1g2rel}
\end{equation}
This relation highlights the unified geometric structure of our framework: Both the mean dissipation and its fluctuations are governed by the same underlying metric tensor, providing a central organizing principle.

Using the metrics $g^{(i)}_{\mu\nu}$ ($i=1, 2$), we can define the thermodynamic length $\mathcal{L}^{(i)}$ of the closed path $\mathcal{C}$:
\begin{align}
  \mathcal{L}^{(i)} \equiv \int_0^\tau dt\, \sqrt{g^{(i)}_{\mu\nu}(t) \dot{\lambda}_\mu(t)\, \dot{\lambda}_\nu(t)} = \oint_{\mathcal{C}} \sqrt{g^{(i)}_{\mu\nu}\, d\lambda_\mu\, d\lambda_\nu}\,.\label{eq:tdlengthi}
\end{align}
Note that $\mathcal{L}^{(i)}$ is a geometric quantity whose value is uniquely determined by the closed path $\mathcal{C}$, independent of the specific driving protocol along the path. 
Applying the Cauchy-Schwarz inequality to Eqs.~(\ref{eq:avra}) and (\ref{eq:vara}), the mean and the variance of $A$ can be lower bounded with $\mathcal{L}^{(i)}$ as
\begin{align}
  \avg{A} \ge \frac{(\mathcal{L}^{(1)})^2}{\tau}\label{eq:bound1}
\end{align}
and
\begin{align}
  \avg{\Delta A^2} \ge \frac{(\mathcal{L}^{(2)})^2}{\tau}\,.\label{eq:bound2}
\end{align}

Following Ref.~\cite{Brandner20}, we define the efficiency $\epsilon$ of the cycle as the ratio of the average work output to the average effective thermal energy input,
\begin{equation}
  \epsilon \equiv \frac{\avg{W}}{\avg{U}}\,.\label{eq:efficiency}
\end{equation}
In the linear-response regime, where dissipation is small, $\epsilon$ can be approximated as
\begin{equation}
  \epsilon \simeq 1 - \frac{\avg{A}}{\mathcal{W}_{\mathrm{qs}}}\,,
\end{equation}
where $\mathcal{W}_{\mathrm{qs}}$ is the work output in the quasistatic limit. Note that $\mathcal{W}_{\mathrm{qs}}$ is deterministic, i.e., $\avg{\mathcal{W}_{\mathrm{qs}}}=\mathcal{W}_{\mathrm{qs}}$~\cite{Sekimotobook10}. From Eq.~(\ref{eq:bound1}), the efficiency $\epsilon$ for a given cycle with path length $\mathcal{L}^{(1)}$ in the $g^{(1)}_{\mu\nu}$ manifold is bounded as~\cite{Brandner20}
\begin{equation}
  \epsilon \le \epsilon_{\mathrm{geo}} \equiv 1 - \frac{(\mathcal{L}^{(1)})^2}{\mathcal{W}_{\mathrm{qs}}\tau}\,.\label{eq:geobound1}
\end{equation}
We also introduce the stochastic efficiency $\mathcal{E}$ defined as~\cite{Watanabe22}
\begin{equation}
  \mathcal{E} \equiv \frac{W}{U}\,.
\end{equation}
In the linear-response regime, $\mathcal{E}$ can be approximated as $\mathcal{E} \simeq 1 - (A/\mathcal{W}_{\mathrm{qs}})$, and its variance is therefore given by
\begin{equation}
  \avg{\Delta \mathcal{E}^2} \simeq \frac{\avg{\Delta A^2}}{\mathcal{W}_{\mathrm{qs}}^2}\,.\label{eq:var_efficiency}
\end{equation}
From Eq.~(\ref{eq:bound2}), the variance $\avg{\Delta \mathcal{E}^2}$ for a given cycle, whose path length in $g^{(2)}_{\mu\nu}$ manifold is $\mathcal{L}^{(2)}$, is bounded as
\begin{equation}
  \avg{\Delta \mathcal{E}^2} \ge \avg{\Delta \mathcal{E}^2}^{\mathrm{geo}} \equiv \frac{(\mathcal{L}^{(2)})^2}{\mathcal{W}_{\mathrm{qs}}^2\tau}\,.\label{eq:geobound2} 
\end{equation}
Note that the bounds on $\epsilon$ and $\avg{\Delta \mathcal{E}^2}$, given by Eqs.~(\ref{eq:geobound1}) and (\ref{eq:geobound2}), are geometric in nature: Their values are fixed irrespective of the protocol once the closed path of the cycle is specified.

From Eqs.~(\ref{eq:geobound1}) and (\ref{eq:geobound2}), we can obtain a geometric lower bound on the relative fluctuation of the stochastic efficiency:
\begin{equation}
  \frac{\sqrt{\avg{\Delta \mathcal{E}^2}}}{\epsilon} \ge \frac{\sqrt{\avg{\Delta \mathcal{E}^2}^{\mathrm{geo}}}}{\epsilon_{\mathrm{geo}}} = \frac{\mathcal{L}^{(2)}}{\sqrt{\tau} \left( \mathcal{W}_{\mathrm{qs}} - \frac{(\mathcal{L}^{(1)})^2}{\tau} \right)}\,.\label{eq:geobound_flucteff}
\end{equation}
This result shows that the relative fluctuation of the stochastic efficiency cannot be made arbitrarily small for a finite cycle duration, but is instead constrained by geometric quantities characterizing the cycle path.

In the following sections, we apply this general framework to three representative systems: (i) $N$-state Markov jump systems (Sec.~\ref{sec:markovjump}), (ii) an overdamped Brownian particle in a 1D power-law potential (Sec.~\ref{sec:overdampedbp}), and (iii) an underdamped Brownian particle in a 1D harmonic oscillator potential (Sec.~\ref{sec:underdampedbp}).

\section{$N$-state Markov jump system}\label{sec:markovjump}

In this section, we consider a continuous-time Markov jump process on an $N$-state classical system, which serves as a minimum model of a classical stochastic system in contact with a thermal environment. The state of the system under the ensemble average is described by the probability $P_i(t)$ of finding the system in state $i$ ($i=1, 2, ... , N$), which satisfies the normalization condition $\sum_i P_i(t) = 1$. The dynamics of this system is governed by the master equation
\begin{equation}
  \frac{d P_i(t)}{dt} = \sum_{j(\ne i)} \left[W_{j\rightarrow i} P_j(t) - W_{i \rightarrow j} P_i(t)\right]\,,\label{eq:mastereq}
\end{equation}
where $W_{k \rightarrow k'}$ denotes the transition rate (i.e., the probability per unit time) from state $k$ to $k'$, which generally depends on the control parameter $\lambda_w$.
Equivalently, Eq.~(\ref{eq:mastereq}) can be written in matrix form as
\begin{equation}
  \dot{\mathbf{P}} = R \mathbf{P}\,,\label{eq:mastereq_matrix}
\end{equation}
where $\mathbf{P} \equiv (P_1, P_2, ... , P_N)^\mathsf{T}$ is the probability vector and the transition-rate matrix $R$ is given by
\begin{equation}
  R \equiv
  \begin{pmatrix}
    -\displaystyle\sum_{j(\neq 1)} W_{1 \rightarrow j} & W_{2 \rightarrow 1} & \cdots & W_{N \rightarrow 1} \\
    W_{1 \rightarrow 2} & -\displaystyle\sum_{j (\neq 2)} W_{2 \rightarrow j} & \cdots & W_{N \rightarrow 2} \\
    \vdots & \vdots & \ddots & \vdots \\
    W_{1 \rightarrow N} & W_{2 \rightarrow N} & \cdots & -\displaystyle\sum_{j (\neq N)} W_{N \rightarrow j}\
  \end{pmatrix}\,.\label{eq:ratematrix}
\end{equation}

When the system is in equilibrium with the thermal environment, it relaxes to the steady state given by the canonical distribution $P_i=\pi_i$,
\begin{equation}
  \pi_i(\lambda_w) = \frac{e^{-\beta H_{\lambda_w}(i)}}{Z_{\lambda_w}}\,,\label{eq:pi_i}
\end{equation}
where $H_{\lambda_w}(i)$ is the energy of state $i$ and $Z_{\lambda_w} \equiv \sum_i e^{-\beta H_{\lambda_w}(i)}$ is the partition function.
To ensure this property, we impose that the transition rates satisfy the detailed-balance condition,
\begin{equation}
  W_{i \rightarrow j} \pi_i = W_{j \rightarrow i} \pi_j\,.\label{eq:db}
\end{equation}

Under this condition, the transition-rate matrix $R$ can be transformed into a symmetric matrix $H$ by a similarity transformation (see, e.g., Ref.~\cite{Risken_book}),
\begin{equation}
  H \equiv S^{-1} R S\,,
\end{equation}
with
\begin{equation}
  S \equiv \operatorname{diag}(\sqrt{\pi_1}, \sqrt{\pi_2},\ldots, \sqrt{\pi_N})\,.
\end{equation}
Consequently, the matrix $R$ is diagonalizable with real eigenvalues $-\Lambda_k$ and a complete set of right and left eigenvectors, denoted by $\phi^{(k)}$ and $\psi^{(k)}$, respectively:
\begin{align}
  R \phi^{(k)} &= -\Lambda_k \phi^{(k)}\,,\\
  (\psi^{(k)})^{\mathsf{T}} R &= -\Lambda_k (\psi^{(k)})^{\mathsf{T}}\,.
\end{align}
These eigenvectors satisfy the biorthonormality condition
\begin{equation}
  (\psi^{(k)})^{\mathsf{T}} \phi^{(l)} = \delta_{k, l}\label{eq:biorthonormal}
\end{equation}
and the completeness relation
\begin{equation}
  \sum_{k=0}^{N-1} \phi^{(k)} (\psi^{(k)})^{\mathsf{T}} = I\,.\label{eq:complete}
\end{equation}
The eigenvalues are ordered as $0=\Lambda_0 < \Lambda_1 \le \Lambda_2 \cdots \le \Lambda_{N-1}$, where the zero mode corresponds to the steady state.

Using the completeness relation (\ref{eq:complete}), the matrix $R$ can be decomposed as
\begin{equation}
  R = -\sum_{k=0}^{N-1}\Lambda_k \phi^{(k)} (\psi^{(k)})^{\mathsf{T}}\,.\label{eq:Rdecomp}
\end{equation}
The propagator $U(t)=e^{Rt}$, defined by $P_i(t) = U(t) P_i(0)$ with $U(0) = I$, is thus
\begin{equation}
  U(t) = e^{Rt} = \sum_{k=0}^{N-1} e^{-\Lambda_k t} \phi^{(k)} (\psi^{(k)})^{\mathsf{T}}\,.
\end{equation}

For quantities $A$ and $B$, represented as column vectors, the equilibrium two-time correlation function reads
\begin{align}
  \avg{A(t)\, B(0)}_{\mathrm{eq}} &= \sum_{n, m} A_n \left[U(t)\right]_{nm} \pi_m B_m\nonumber\\
  &= \sum_{k=0}^{N-1} e^{-\Lambda_k t} \sum_n A_n \phi^{(k)}_n \sum_m B_m \psi^{(k)}_m \pi_m\,.\label{eq:corrfunc_ab}
\end{align}

\subsection{Example: classical two-level system}

For the sake of illustration, we now focus on the simplest nontrivial case: a classical two-level system with states $\sigma = \pm 1$, whose Hamiltonian is given by
\begin{equation}
  H(\sigma) = \frac{\Delta}{2} \sigma\,.\label{eq:h_tls}
\end{equation}
Here, we identify the energy gap $\Delta$ as a mechanical control parameter $\lambda_w = \Delta$. Introducing the probability vector $\mathbf{P} \equiv (P_+, P_-)^\mathsf{T}$ and the shorthand notation of the transition rates $r_+ \equiv W_{- \rightarrow +}$ and $r_- \equiv W_{+ \rightarrow -}$, the transition-rate matrix $R$ can be written as
\begin{equation}
  R =
  \begin{pmatrix}
    -r_- & r_+ \\
    r_- & -r_+
  \end{pmatrix}\,.\label{eq:ratematrix_tls}
\end{equation}
For the transition rates, we adopt the symmetric choice
\begin{align}
  r_+ = \gamma e^{-\beta\Delta/2} \quad\mbox{and}\quad r_- = \gamma e^{\beta\Delta/2}\,,
\end{align}
which satisfies the detailed-balance condition,
\begin{equation}
  \frac{r_+}{r_-} = e^{-\beta\Delta}\,.\label{eq:db_tls}
\end{equation}
Here, the positive coefficient $\gamma$ sets the overall rate scale.

With this choice of the transition rates, eigenvalues of $R$ are
\begin{align}
  \Lambda_0 &= 0\,,\label{eq:lambda0_tls}\\
  \Lambda_1 &= r_+ + r_- = 2 \gamma\, \cosh{\left(\frac{\beta\Delta}{2}\right)}\,.\label{eq:lambda1_tls}
\end{align}

The steady-state distribution $P_\pm=\pi_\pm$, corresponding to the zero mode $\Lambda_0=0$, is given by the canonical distribution associated with the Hamiltonian~(\ref{eq:h_tls}):
\begin{equation}
  \pi_\pm = \frac{r_\pm}{r_+ + r_-} = \frac{e^{\mp\beta\Delta/2}}{2 \cosh{\left(\dfrac{\beta\Delta}{2}\right)}}\,.\label{eq:pi_pm}
\end{equation}
The equilibrium magnetization $m$ is given by
\begin{equation}
  m \equiv \avg{\sigma}_{\mathrm{eq}} = \pi_+ - \pi_- = -\tanh{\frac{\beta\Delta}{2}}.\label{eq:m}
\end{equation}
Using the spectral decomposition of the transition-rate matrix $R$, the equilibrium two-time correlation function of $\sigma$ can be evaluated analytically (see Appendix~\ref{app:corrfunc} for details), yielding
\begin{align}
  \avg{\sigma(t)\, \sigma(0)}_{\mathrm{eq}} = m^2 + (1-m^2) e^{-\Lambda_1 t}\,.\label{eq:corrfunc_sigma}
\end{align}

Using this result, we derive the equilibrium two-time correlation functions $\avg{\Delta X_\mu(t) \Delta X_\nu(0)}_{\mathrm{eq}}$ of the fluctuations of the generalized forces to determine the metrics of the thermodynamic length. The generalized forces conjugate to the control parameters $\lambda_w = \Delta$ and $\lambda_u = T$ are given by
\begin{align}
  X_w &= -\frac{\partial H(\sigma)}{\partial \Delta} = -\frac{\sigma}{2}\,,\label{eq:xw_tls}\\
  X_u &= S(\sigma) = -k_{\mathrm B} \ln{P_\sigma}\,,\label{eq:xu_tls}
\end{align}
respectively. Under the equilibrium ensemble average considered here, $X_u$ reduces to
\begin{equation}
  X_u =  -k_{\mathrm B} \ln{\pi_\sigma}\,.
\end{equation}
Using the result of Eq.~(\ref{eq:corrfunc_sigma}), the equilibrium two-time correlation functions $\avg{\Delta X_\mu(t) \Delta X_\nu(0)}_{\mathrm{eq}}$ of the fluctuations of the generalized forces follow (see Appendix~\ref{app:corrfunc}):
\begin{equation}
  \avg{\Delta X_\mu(t) \Delta X_\nu(0)}_{\mathrm{eq}} \propto (1-m^2) e^{-\Lambda_1 t}\,,
\end{equation}
leading to the correlation times $\tau_{\mu\nu}$ and weights $C_{\mu\nu}$:
\begin{align}
  \tau_{ww} &= \tau_{uu} = \tau_{wu} = \tau_{uw} = 1/\Lambda_1\,,\\
  C_{ww} &= C_{uu} = C_{wu} = C_{uw} = 1\,,
\end{align}
and the covariance tensor $\sigma_{\mu\nu}$:
\begin{align}
  \sigma_{ww} &= \frac{1}{4} (1-m^2)\,,\\
  \sigma_{uu} &= k_{\mathrm{B}}^2 \frac{\beta^2\Delta^2}{4} (1-m^2)\,,\\
  \sigma_{wu} &= \sigma_{uw} = -k_{\mathrm{B}}\frac{\beta\Delta}{4} (1-m^2)\,.
\end{align}
As a result, from Eq.~(\ref{eq:g1munu}), we finally obtain the metric tensor $g^{(1)}_{\mu\nu}$ as
\begin{equation}
  g^{(1)}_{\mu\nu} = \beta \frac{1-m^2}{4\Lambda_1}
  \begin{bmatrix}
    1 & -\dfrac{\Delta}{T}\medskip\\
    -\dfrac{\Delta}{T} \quad& \left(\dfrac{\Delta}{T}\right)^2
  \end{bmatrix}\,.\label{eq:g1munu_tls}
\end{equation}
The metric tensor $g^{(2)}_{\mu\nu}$ for the variance is then immediately given by Eq.~(\ref{eq:g1g2rel}).

From Eq.~(\ref{eq:g1munu_tls}), one can readily see that the determinants of the metric tensors $g^{(1)}_{\mu\nu}$ and $g^{(2)}_{\mu\nu} = 2k_{\mathrm{B}}T g^{(1)}_{\mu\nu}$ vanish; hence, both metrics are singular with a zero eigenvalue. This singularity of the metric can be understood directly from the equilibrium distribution $\pi_{\pm} \propto e^{\mp \beta\Delta/2}$ given by Eq.~(\ref{eq:pi_pm}), which depends on the control parameters only through the combination $\Delta/T$. Consequently, there exists a direction in the parameter space along which the equilibrium distribution remains unchanged, giving rise to the zero eigenvalue of the thermodynamic metric. The normalized eigenvector $\mathbf{v}_0$ corresponding to this zero eigenvalue is $\mathbf{v}_0 = [(\Delta/T)^2 + 1]^{-1/2} (\Delta/T, 1)^\intercal$, whose direction is given by $dT/d\Delta = T/\Delta$. This direction corresponds to straight lines on the $T$-$\Delta$ plane connecting each point $(\Delta, T)$ to the origin. A path along the zero-eigenvalue direction of the metrics $g^{(i)}_{\mu\nu}$, given by $T/\Delta = \mbox{constant}$, therefore corresponds to an isentropic path along which the equilibrium entropy remains constant, which is given by
\begin{align}
  \avg{S}_{\mathrm{eq}} &= -k_{\mathrm{B}} \avg{\ln{\pi_\sigma}}_{\mathrm{eq}}\nonumber\\
  &= k_{\mathrm{B}} \left[ \ln{\left(2 \cosh{\frac{\beta\Delta}{2}} \right)} + \frac{\beta\Delta}{2} m(\beta\Delta) \right].
\end{align}
Consequently, along isentropic strokes, both the average and the variance of $A$ vanish, $\avg{A} = \avg{\Delta A^2} = 0$, within the linear-response regime.

\section{Overdamped Brownian particle in a 1D power-law potential}\label{sec:overdampedbp}

In this section, we consider an overdamped Brownian particle in a class of 1D power-law potentials of the form
\begin{align}
  V_{\lambda_w}(x) = \frac{\lambda_w^{\alpha}}{2n} x^{2n}\,,\label{eq:powerlawpot}
\end{align}
where $n \ge 1$ is a positive integer and $\alpha=1$ or $-2n$.
The exponent $\alpha$ specifies how the control parameter $\lambda_w$ enters the potential: $\alpha=1$ corresponds to a strength-controlled potential, and $\alpha=-2n$ corresponds to a width-controlled potential.

As we show below, unlike the example in the previous section, the equilibrium correlation functions $\avg{\Delta X_\mu(t)\, \Delta X_\nu(t')}_{\mathrm{eq}}$ for $n \ne 1$ are given by an infinite sum of exponentials in time with different relaxation timescales $\tau_{\mu\nu}^{(i)}$ ($i=1, 2, 3, ...$). Here, all $\tau_{\mu\nu}^{(i)}$ share the same dependence on the friction coefficient $\gamma$, temperature $T$, and mechanical control parameter $\lambda_w$ as $\tau_{\mu\nu}^{(i)} \propto \gamma T^{-1 + \frac{1}{n}} \lambda_w^{-\frac{\alpha}{n}}$, while the corresponding coefficients $C_{\mu\nu}^{(i)}$ are independent of $\gamma$, $T$, and $\lambda_w$. Consequently, the effective correlation time $\overline{\tau}_{\mu\nu}$ also scales as
\begin{align}
  \overline{\tau}_{\mu\nu} \propto \gamma T^{-1 + \frac{1}{n}} \lambda_w^{-\frac{\alpha}{n}}\,,
\end{align}
and the proportionality constant for each value of $n$ can be determined numerically.

\subsection{Equilibrium state and generalized forces}

The equilibrium state $\rho_{\mathrm{eq}}(x)$ of the overdamped particle in the potential (\ref{eq:powerlawpot}) is given by the canonical distribution
\begin{equation}
  \rho_{\mathrm{eq}}(x) = Z^{-1} \exp{\left[-\beta \frac{\lambda_w^\alpha}{2n} x^{2n}\right]}\,,\label{eq:rhoeq}
\end{equation}
with the partition function
\begin{equation}
  Z = \int_{-\infty}^{\infty}dx\, e^{-\beta V_{\lambda_w}(x)} = \frac{1}{n}\Gamma\left(\frac{1}{2n}\right) \left(\frac{\beta}{2n}\right)^{-\frac{1}{2n}} \lambda_w^{-\frac{\alpha}{2n}}\,.\label{eq:z_overdampedbp}
\end{equation}
Here, $\Gamma(x) \equiv \int_0^{\infty} z^{x-1} e^{-z} dz$ denotes the Gamma function.

The generalized forces $X_\mu$ conjugate to the control parameters $\lambda_\mu$ are given by
\begin{align}
  X_w &= -\frac{\partial V_{\lambda_w}}{\partial \lambda_w} = -\alpha \frac{\lambda_w^{\alpha-1}}{2n} x^{2n}\,,\\
  X_u &= -k_{\mathrm{B}} \ln{\rho_{\mathrm{eq}}} = k_{\mathrm{B}} \ln{Z} + \frac{\lambda_w^\alpha}{2n T} x^{2n}\,. 
\end{align}
Here, we have used the equilibrium state $\rho_{\mathrm{eq}}$ to define $X_u$ since we always consider equilibrium ensemble averages, as in the previous section. Because both $\Delta X_w$ and $\Delta X_u$ are proportional to $\Delta(x^{2n}) \equiv x^{2n} - \avg{x^{2n}}$, the equilibrium two-time correlation functions $\avg{\Delta X_\mu(t)\, \Delta X_\nu(0)}_{\mathrm{eq}}$ for any $\mu$ and $\nu$ are proportional to the two-time correlation function $C_{x^{2n}}(t)$ of $\Delta (x^{2n})$, i.e.,
\begin{equation}
  \avg{\Delta X_\mu(t)\, \Delta X_\nu(0)}_{\mathrm{eq}} \propto C_{x^{2n}}(t)\,,
\end{equation}
where
\begin{align}
  C_{x^{2n}}(t) &\equiv \avg{\Delta (x^{2n}(t))\, \Delta (x^{2n}(0))}_{\mathrm{eq}}\nonumber\\
  &= \avg{x^{2n}(t)\, x^{2n}(0)}_{\mathrm{eq}} - \avg{x^{2n}}_{\mathrm{eq}}^2\,.
\end{align}
Thus, the essential quantity in the correlation functions $\avg{\Delta X_\mu(t)\, \Delta X_\nu(0)}_{\mathrm{eq}}$ is $\avg{x^{2n}(t)\, x^{2n}(0)}_{\mathrm{eq}}$, which is common to all $\mu$ and $\nu$.

Since $C_{x^{2n}}(t)$ is the only time-dependent part of the correlation function $\avg{\Delta X_\mu(t)\, \Delta X_\nu(0)}_{\mathrm{eq}}$ for any $\mu$ and $\nu$, all the correlation times are equal:
\begin{equation}
  \overline{\tau} \equiv \overline{\tau}_{ww} = \overline{\tau}_{wu} = \overline{\tau}_{uw} = \overline{\tau}_{uu}\,.
\end{equation}
The equilibrium correlation functions can therefore be written as
\begin{equation}
  \avg{\Delta X_\mu(t)\, \Delta X_\nu(0)}_{\mathrm{eq}} = \avg{\Delta X_\mu(0)\, \Delta X_\nu(0)}_{\mathrm{eq}}\, \frac{C_{x^{2n}}(t)}{C_{x^{2n}}(0)}\,.\label{eq:corrfunc_overdampedbp}
\end{equation}

\subsection{Spectral representation of the correlation function}

We now calculate $\avg{x^{2n}(t)\, x^{2n}(0)}_{\mathrm{eq}}$. This quantity can be expressed in terms of the transition probability $\rho(x,t|x',0)$ as
\begin{align}
  \avg{x^{2n}(t)\, x^{2n}(0)}_{\rm eq} = \int dx \int dx'\, x^{2n}\, x'^{2n}\, \rho(x,t|x',0) \rho_{\mathrm{eq}}(x')\,.\label{eq:x2nx2n}
\end{align}
Hence, the problem reduces to computing the transition probability $\rho(x,t|x',0)$. This can be achieved using a spectral decomposition of the evolution operator (see Appendix~\ref{app:decomp_fpop} and Ref.~\cite{Risken_book}). The dynamics of the overdamped Brownian particle is governed by the Fokker-Planck equation
\begin{equation}
  \frac{\partial}{\partial t} \rho(x, t) = L_{\mathrm{FP}} \rho(x, t)\,,\label{eq:FPeq}
\end{equation}
with the Fokker-Planck operator $L_{\mathrm{FP}}$ defined as
\begin{equation}
  L_{\mathrm{FP}} \equiv \frac{\partial}{\partial x}\left[\frac{1}{\gamma} \frac{\partial V_{\lambda_w}}{\partial x} + \frac{k_{\mathrm{B}}T}{\gamma} \frac{\partial}{\partial x} \right]\,.\label{eq:FPop}
\end{equation}
Let $\{\Lambda_i,\, \phi_i\}$ denote the eigenvalues and eigenfunctions of $L_{\mathrm{FP}}$,
\begin{equation}
  L_{\mathrm{FP}} \phi_i = -\Lambda_i \phi_i\,,\label{eq:FPeigenvalueeq}
\end{equation}
where the eigenvalues are sorted in an ascending order with $i$ as $0 = \Lambda_0 < \Lambda_1 \le \Lambda_2 \le \cdots$. The zero mode $\phi_0$ with $\Lambda_0 = 0$ corresponds to the steady state up to a normalization constant: $\rho_{\mathrm{eq}} \propto \phi_0$.
The completeness relation reads
\begin{equation}
  \delta(x-x') = \sum_{i=0}^\infty \tilde{\phi}_i(x)\, \tilde{\phi}_i(x')\,,\label{eq:complete_fp}
\end{equation}
where $\tilde{\phi}_i$ are the similarity-transformed eigenfunctions defined as~\cite{note:phitilde}
\begin{align}
  \tilde{\phi}_i \equiv e^{\Phi/2}\phi_i\,,
\end{align}
with~\cite{note:Phi}
\begin{align}
  \Phi(x) \equiv \ln{\left(\frac{k_{\mathrm{B}}T}{\gamma}\right)} + \beta V_{\lambda_w}(x)\,.
\end{align}
With this definition, the equilibrium state is given by
\begin{equation}
  \rho_{\mathrm{eq}}(x) = \tilde{\phi}_0^2(x)\,.
\end{equation}
Using the completeness relation (\ref{eq:complete_fp}), the transition probability can be expanded as
\begin{align}
  \rho(x,t|x',t') &= e^{(t-t')L_{\mathrm{FP}}(x)} \delta(x-x')\nonumber\\
  &= \frac{\tilde{\phi}_0(x)}{\tilde{\phi}_0(x')} \sum_{i=0}^\infty e^{-\Lambda_i (t-t')} \tilde{\phi}_i(x)\, \tilde{\phi}_i(x')\,.\label{eq:transition_prob}
\end{align}
Substituting this expression into Eq.~(\ref{eq:x2nx2n}), we finally obtain the eigenfunction expansion of the correlation function:
\begin{align}
  &\avg{x^{2n}(t)\, x^{2n}(0)}_{\rm eq}\nonumber\\
  =& \int dx \int dx'\, x^{2n}\, x'^{2n}\, \tilde{\phi}_0(x)\, \tilde{\phi}_0(x') \sum_{i=0}^\infty e^{-\Lambda_i t} \tilde{\phi}_i(x)\, \tilde{\phi}_i(x')\nonumber\\
  =& \sum_{i=0}^\infty c_i\, e^{-\Lambda_i\, t}\,,
\end{align}
with
\begin{equation}
  c_i \equiv \left(\int_{-\infty}^{\infty} dx\, x^{2n}\, \tilde{\phi}_0(x)\, \tilde{\phi}_i(x)\right)^2\,.\label{eq:ci}
\end{equation}
Since $c_0 = \left(\int dx\, x^{2n} \rho_{\mathrm{eq}}(x)\right)^2 = \avg{x^{2n}}_{\mathrm{eq}}^2$, the correlation function $C_{x^{2n}}$ can be written as
\begin{equation}
  C_{x^{2n}}(t) = \sum_{i=1}^{\infty} c_i e^{-\Lambda_i t}\,,
\end{equation}
where the sum starts from $i=1$.

With this expression of $C_{x^{2n}}$, Eq.~(\ref{eq:corrfunc_overdampedbp}) can be rewritten as
\begin{equation}
  \avg{\Delta X_\mu(t)\, \Delta X_\nu(0)}_{\mathrm{eq}}
  = \sigma_{\mu\nu} \sum_{i=1}^\infty \tilde{c}_i\, e^{-\Lambda_i t}\,,
\end{equation}
where
\begin{equation}
  \tilde{c}_i \equiv \dfrac{c_i}{\sum_{j=1}^\infty c_j}\,.
\end{equation}

\subsection{Scaling of eigenvalues and effective correlation time}

Now, we identify the scaling of the eigenvalues $\Lambda_i$. Introducing the dimensionless coordinate $z \equiv a x$ with
\begin{equation}
  a \equiv \left(\frac{\lambda_w^\alpha}{2n k_{\mathrm{B}}T} \right)^{\frac{1}{2n}}\,,
\end{equation}
the eigenvalue equation (\ref{eq:FPeigenvalueeq}) of $L_{\mathrm{FP}}$ can be rewritten as a Sturm-Liouville problem,
\begin{equation}
  \frac{d}{dz} \left[e^{-z^{2n}} \frac{d}{dz} y_i(z)\right] = -\left[ a^{-2} \frac{\gamma}{k_{\mathrm{B}}T} \Lambda_i\right] e^{-z^{2n}} y_i(z)\,,\label{eq:dimensionlessFP}
\end{equation}
where $y_i \equiv e^{\Phi/2}\tilde{\phi}_i = e^{\Phi} \phi_i$. Since the factor inside the brackets in the right-hand side of Eq.~(\ref{eq:dimensionlessFP}) is dimensionless, $\Lambda_i$ should scale as
\begin{align}
  \Lambda_i \propto \frac{k_{\mathrm{B}}T}{\gamma}a^2 \propto \gamma^{-1} (k_{\mathrm{B}}T)^{1-\frac{1}{n}} \lambda_w^{\frac{\alpha}{n}}\,.\label{eq:scaling}
\end{align}

\begin{figure}[t!]
\centering
\includegraphics[width=0.93 \columnwidth]{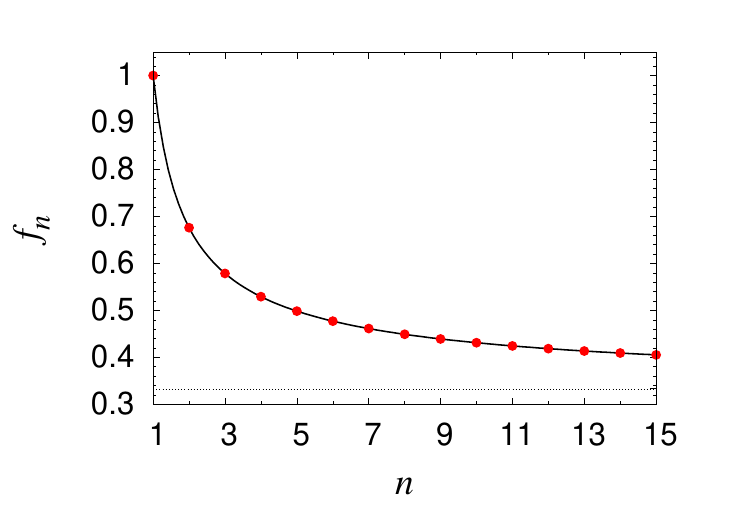}
\caption{Prefactor $f_n$ as a function of $n$. The horizontal dotted line indicates the asymptotic value $f_{n\rightarrow \infty} = 1/3$. The black solid line shows the analytical expression given by Eq.~(\ref{eq:fn}), while the red dots represent the numerical results obtained by solving the Sturm-Liouville problem.}
\label{fig:fn}
\end{figure}

Solving the Sturm-Liouville problem (\ref{eq:dimensionlessFP}) to determine $\{\Lambda_i,\, \tilde{\phi}_i\}$ (details are given in Appendix~\ref{app:slproblem}), together with the scaling of $\Lambda_i$ given by Eq.~(\ref{eq:scaling}), the effective correlation times $\overline{\tau}_{\mu\nu}$ are obtained as
\begin{align}
  \overline{\tau}_{\mu\nu} &= \sum_{i=1}^\infty \frac{\tilde{c}_i}{\Lambda_i}\nonumber\\
  &=\frac{f_n}{2n} \gamma\, (k_{\mathrm{B}}T)^{-1+\frac{1}{n}} \lambda_w^{-\frac{\alpha}{n}}\,.\label{eq:taubar_overdampedbp}
\end{align}
Here, $f_n$ is a dimensionless function depending only on $n$, which is given analytically by
\begin{equation}
  f_n = (2n)^{\frac{1}{n}}\, \frac{\Gamma\left(\dfrac{3}{2n}\right)}{\Gamma\left(\dfrac{1}{2n}\right)}\,.\label{eq:fn}
\end{equation}
The derivation of Eq.~(\ref{eq:fn}) is presented in Appendix~\ref{app:fn}. The function $f_n$ decreases monotonically with $n$ starting from $f_{n=1}=1$ (see Fig.~\ref{fig:fn}). For large $n$, it decreases relatively slowly as $\sim \ln{n}/n$ to a nonzero asymptotic value, $f_{n\rightarrow\infty}=1/3$.

\subsection{Metric tensors}

For the equilibrium state (\ref{eq:rhoeq}), the covariance tensor $\sigma_{\mu\nu}$ is given by
\begin{align}
  \sigma_{ww} &= \frac{\alpha^2}{2n} \frac{1}{\beta^2 \lambda_w^2}\,,\\
  \sigma_{uu} &= \frac{1}{2n} k_{\mathrm{B}}^2\,,\\
  \sigma_{wu} &= \sigma_{uw} = -\frac{\alpha}{2n} k_{\mathrm{B}}  \frac{1}{\beta \lambda_w}\,.\label{eq:sigmawu_overdampedbp}
\end{align}
Substituting Eqs.~(\ref{eq:taubar_overdampedbp})--(\ref{eq:sigmawu_overdampedbp}) into Eq.~(\ref{eq:g1munu}), we finally obtain the metric tensor $g_{\mu\nu}^{(1)}$ for $\avg{A}$ as
\begin{equation}
  g^{(1)}_{\mu\nu} = \frac{\alpha^2\, f_n}{(2n)^2} \gamma\, (k_{\mathrm{B}}T)^{\frac{1}{n}}\, \lambda_w^{-2-\frac{\alpha}{n}}
  \begin{bmatrix}
    1 & -\dfrac{\lambda_w}{\alpha T}\medskip\\
    -\dfrac{\lambda_w}{\alpha T} \quad& \left(\dfrac{\lambda_w}{\alpha T}\right)^2
  \end{bmatrix}\,.\label{eq:g1munu_odb}
\end{equation}
The metric tensor $g^{(2)}_{\mu\nu}$ for the variance follows immediately from Eq.~(\ref{eq:g1g2rel}).

As in the example of the previous section [Eq.~(\ref{eq:g1munu_tls})], the metric tensors $g_{\mu\nu}^{(i)}$ in the present case are also singular and possess a zero eigenvalue. Again, this singularity can be understood directly from the equilibrium distribution, $\rho_{\mathrm{eq}}(x) \propto \exp{\left[-\beta \lambda_w^\alpha x^{2n}/(2n)\right]}$, which depends on the control parameters only through the combination $\lambda_w^\alpha/T$. Consequently, there exists a direction in the parameter space along which the equilibrium distribution remains unchanged, giving rise to a zero-eigenvalue direction of the metric. The eigenvector $\mathbf{v}_0$ corresponding to this zero eigenvalue is given by $\mathbf{v}_0 \propto (\lambda_w/\alpha T,\, 1)^\intercal$. Accordingly, the direction along this eigenvector satisfies $dT/d\lambda_w = \alpha T/\lambda_w$, so that the path along the zero eigenvalue is given by $T/\lambda_w^{\alpha} = \mbox{constant}$. Along this path, the equilibrium entropy
\begin{align}
  \avg{S}_{\mathrm{eq}} =&\, k_{\mathrm{B}} \bigg\{ \ln{\left[\Gamma\left(\frac{1}{2n}\right)\right]} - \ln{n} + \frac{1}{2n}(1+\ln{2n}) \nonumber\\
   &- \frac{1}{2n} \ln{\left(\beta\lambda_w^\alpha\right)} \bigg\}
\end{align}
remains constant. Consequently, as in the previous section, both the mean and the variance of the dissipated availability vanish along isentropic strokes, $\avg{A} = \avg{\Delta A^2} = 0$, within the linear-response approximation.

The other eigenvalue $\lambda_1$ of the metric tensor (\ref{eq:g1munu_odb}) is given by
\begin{equation}
  \lambda_1 = \frac{\alpha^2\, f_n}{(2n)^2} \gamma\, (k_{\mathrm{B}}T)^{\frac{1}{n}}\, \lambda_w^{-2-\frac{\alpha}{n}} \left[1+\left(\frac{\lambda_w}{\alpha T}\right)^2\right]\,.
\end{equation}
This nonzero eigenvalue gives the spectral norm of $g^{(1)}_{\mu\nu}$, $||g^{(1)}_{\mu\nu}||_2 = \lambda_1$ and thus characterizes the overall magnitude of the metric. For $\alpha=1$ (strength-controlled case), $\lambda_1$ reduces to $\lambda_1 = \frac{f_n}{(2n)^2} (k_{\mathrm{B}}T/\lambda_w)^{\frac{1}{n}} \gamma\, \lambda_w^{-2} [1+(\lambda_w/T)^2]$. Its $n$-dependent numerical prefactor $f_n/(2n)^2$ decreases rapidly to zero as $n$ increases, starting from $f_1/2^2 = 1/4$ at $n=1$. For $\alpha=-2n$ (width-controlled case), on the other hand, $\lambda_1$ reads $\lambda_1 = f_n (k_{\mathrm{B}}T)^{\frac{1}{n}} \gamma\, [1+(\lambda_w/2nT)^2]$. Here, the numerical prefactor $f_n$ decreases monotonically but relatively slowly with $n$, approaching the nonzero asymptotic value $1/3$ (see Fig.~\ref{fig:fn}). These $n$ dependences indicate that both the mean and the variance of dissipation are reduced for steeper potentials with larger $n$, as long as the linear-response approximation remains valid. This reduction is more pronounced for strength-controlled potentials than for width-controlled ones.

\section{Underdamped Brownian particle in a harmonic potential}\label{sec:underdampedbp}

In this section, we consider an underdamped Brownian particle in a 1D harmonic potential given by Eq.~(\ref{eq:vho}). At the microscopic level, its dynamics is governed by the following Langevin equation:
\begin{equation}
  M\ddot{x} + \lambda_w x + \gamma \dot{x} = \xi(t)\,,\label{eq:langevin}
\end{equation}
where $M$ is the mass of the Brownian particle, $\gamma$ is the friction coefficient, and $\xi(t)$ is a Gaussian white noise whose strength is determined by the temperature $T$ of the thermal environment through the fluctuation-dissipation relation.
The equilibrium two-time correlation functions of the generalized forces for this system have been obtained analytically by Frim and DeWeese~\cite{Frim22}. Here, we take their results as a starting point and focus on the implications for the metric tensor.

For later convenience, we introduce the characteristic relaxation rates
\begin{align}
  \Gamma_\pm
  \equiv \frac{\gamma}{2M} \left[1 \pm \sqrt{1 - \frac{4M\lambda_w}{\gamma^2}}\,\right]\,,\label{eq:gamma_pm}
\end{align}
which naturally emerge from the linear underdamped Langevin dynamics. As shown below, using the rates $\Gamma_\pm$, the equilibrium two-time correlation functions can be expressed as a superposition of three exponential relaxation modes, characterized by the time constants
\begin{align}
  \tau_{\mu\nu}^{(1)} =& (2\Gamma_-)^{-1}\,,\label{eq:tau1_udb}\\
  \tau_{\mu\nu}^{(2)} =& (\Gamma_- + \Gamma_+)^{-1}\,,\label{eq:tau2_udb}\\
  \tau_{\mu\nu}^{(3)} =& (2\Gamma_+)^{-1}\,.\label{eq:tau3_udb}
\end{align}
For sufficiently weak damping, $\gamma^2 < 4M\lambda_w$, the rates $\Gamma_\pm$ form a complex-conjugate pair, corresponding to damped oscillatory relaxation. Importantly, even in this case, the correlation functions retain the same formal decomposition into three exponential modes, and the expressions of the time constants $\tau^{(i)}_{\mu\nu}$ remain valid.

The equilibrium two-time correlation functions $\avg{\Delta X_\mu(t) \Delta X_\nu(0)}_{\rm eq}$ $(t\ge 0)$ are given by~\cite{Frim22}
\begin{widetext}
\begin{align}
  \avg{\Delta X_w(t)\, \Delta X_w(0)}_{\mathrm{eq}} =& \frac{1}{2}\left(\frac{k_{\mathrm{B}}T}{\lambda_w}\right)^2\, \frac{1}{(\Gamma_+ - \Gamma_-)^2} \left( \Gamma_+ e^{-\Gamma_- t} - \Gamma_- e^{-\Gamma_+ t} \right)^2\,,\label{eq:dxwdxw_udb}\\
  \avg{\Delta X_w(t)\, \Delta X_u(0)}_{\mathrm{eq}} =& -\frac{k_{\mathrm{B}}^2 T}{2\lambda_w}\, \frac{1}{(\Gamma_+ - \Gamma_-)^2} \left[ \left(\Gamma_+ e^{-\Gamma_- t} - \Gamma_- e^{-\Gamma_+ t}\right)^2 + \frac{\lambda_w}{M} \left(e^{-\Gamma_- t} - e^{-\Gamma_+ t}\right)^2\right]\,,\label{eq:dxwdxu_udb}\\
  \avg{\Delta X_u(t)\, \Delta X_w(0)}_{\mathrm{eq}} =& -\frac{k_{\mathrm{B}}^2 T}{2\lambda_w}\, \frac{1}{(\Gamma_+ - \Gamma_-)^2} \left[ \left(\Gamma_+ e^{-\Gamma_- t} - \Gamma_- e^{-\Gamma_+ t}\right)^2 + \frac{M}{\lambda_w} (\Gamma_+ \Gamma_-)^2 \left(e^{-\Gamma_-t} - e^{-\Gamma_+t}\right)^2 \right]\,,\label{eq:dxudxw_udb}\\
  \avg{\Delta X_u(t)\, \Delta X_u(0)}_{\mathrm{eq}} =& k_{\mathrm{B}}^2\, \frac{1}{2(\Gamma_+ - \Gamma_-)^2} \left[ \left(\Gamma_+ e^{-\Gamma_- t} - \Gamma_- e^{-\Gamma_+ t}\right)^2 + \left(\Gamma_- e^{-\Gamma_- t} - \Gamma_+ e^{-\Gamma_+ t}\right)^2 + \frac{\lambda_w}{M} \left(e^{-\Gamma_- t} - e^{-\Gamma_+ t}\right)^2 \right.\nonumber\\
    &\left. + \frac{M}{\lambda_w} (\Gamma_+ \Gamma_-)^2 \left(e^{-\Gamma_-t} - e^{-\Gamma_+t}\right)^2 \right]\,.\label{eq:dxudxu_udb}
\end{align}
\end{widetext}
These expressions admit a representation in terms of three exponential modes characterized by the time constants $\tau^{(i)}_{\mu\nu}$ defined above, which is consistent with the general multiexponential structure introduced in Eq.~(\ref{eq:corrmulti}). Explicitly, we can write
\begin{align}
  \avg{\Delta X_\mu(t)\, \Delta X_\nu(t')}_{\rm eq}
  = \sigma_{\mu\nu}(t)\, \sum_{i=1}^{3} C^{(i)}_{\mu\nu}\, e^{-|t-t'|/\tau_{\mu\nu}^{(i)}}\,.\label{eq:correxp_udb}
\end{align}
Here, the explicit expressions of the coefficients $C^{(i)}_{\mu\nu}$ for all tensor components are rather lengthy and are therefore summarized in Appendix~\ref{app:udb}. With these results of $\tau^{(i)}_{\mu\nu}$ and $C^{(i)}_{\mu\nu}$, the effective correlation times $\overline{\tau}_{\mu\nu} = \sum_{i=1}^3 C_{\mu\nu}^{(i)} \tau_{\mu\nu}^{(i)}$ are obtained as
\begin{align}
  \overline{\tau}_{ww} &= \frac{\gamma}{2 \lambda_w} \left( 1 + \frac{M\lambda_w}{\gamma^2} \right)\,,\label{eq:tau_ww}\\
  \overline{\tau}_{wu} = \overline{\tau}_{uw} &= \frac{\gamma}{2 \lambda_w} \left( 1 + 2 \frac{M\lambda_w}{\gamma^2} \right)\,,\label{eq:tau_wu}\\
  \overline{\tau}_{uu} &= \frac{\gamma}{4 \lambda_w} \left( 1 + 4 \frac{M\lambda_w}{\gamma^2} \right)\,.\label{eq:tau_uu}
\end{align}
The remaining ingredient entering the metric tensor is the covariance tensor $\sigma_{\mu\nu}$, which can readily be obtained as
\begin{align}
  &\sigma_{ww} = \frac{1}{2}\left(\frac{k_{\mathrm{B}}T}{\lambda_w}\right)^2\,,\label{eq:sigma_ww}\\
  &\sigma_{wu} = \sigma_{uw} = -\frac{k_{\mathrm{B}}^2 T}{2\lambda_w}\,,\label{eq:sigma_wu}\\
  &\sigma_{uu} = k_{\mathrm{B}}^2\,.\label{eq:sigma_uu}
\end{align}
Substituting the above $\sigma_{\mu\nu}$ and $\overline{\tau}_{\mu\nu}$ into Eq.~(\ref{eq:g1munu_final}), we finally obtain the metric tensor
\begin{equation}
  g^{(1)}_{\mu\nu} = k_{\mathrm{B}}T \frac{\gamma}{4 \lambda_w^3}
  \begin{bmatrix}
     1 + \dfrac{M\lambda_w}{\gamma^2} & -\dfrac{\lambda_w}{T} \left( 1 + 2 \dfrac{M\lambda_w}{\gamma^2} \right)\medskip\\
    -\dfrac{\lambda_w}{T} \left( 1 + 2 \dfrac{M\lambda_w}{\gamma^2} \right) & \dfrac{\lambda_w^2}{T^2} \left( 1 + 4 \dfrac{M\lambda_w}{\gamma^2} \right)\\
  \end{bmatrix}\,.\label{eq:g1munu_udb}
\end{equation}
Same as in the previous sections, the metric tensor $g^{(2)}_{\mu\nu}$ is obtained from Eq.~(\ref{eq:g1g2rel}) using the metric tensor $g^{(1)}_{\mu\nu}$ given above.

The terms proportional to $M\lambda_w/\gamma^2$, which contain the mass $M$ of the particle, originate from inertial effects that are neglected in the overdamped case discussed in the previous section. Due to the presence of these terms, the metric tensors $g^{(i)}_{\mu\nu}$ in the present case are regular, i.e., their determinants are nonzero, and hence all their eigenvalues are nonvanishing, in contrast to the examples discussed in the previous sections. This regularity can also be understood directly from the equilibrium distribution $\rho_{\mathrm{eq}}(x, p) \propto \exp{\left[-\beta \left(\frac{p^2}{2M} + \frac{\lambda_w}{2}x^2 \right)\right]}$. Unlike in the overdamped case, the kinetic-energy contribution introduces a temperature dependence that cannot be absorbed into the combination $\lambda_w/T$. Consequently, no direction in the parameter space leaves the equilibrium distribution unchanged, and the metric no longer has a zero-eigenvalue direction. One can also readily see that, in the overdamped limit $M\lambda_w/\gamma^2 \ll 1$, Eq.~(\ref{eq:g1munu_udb}) consistently reduces to Eq.~(\ref{eq:g1munu_odb}) for $n=1$ and $\alpha=1$, corresponding to the strength-controlled harmonic potential.

\begin{figure*}[t]
\centering
\includegraphics[width=1.69 \columnwidth]{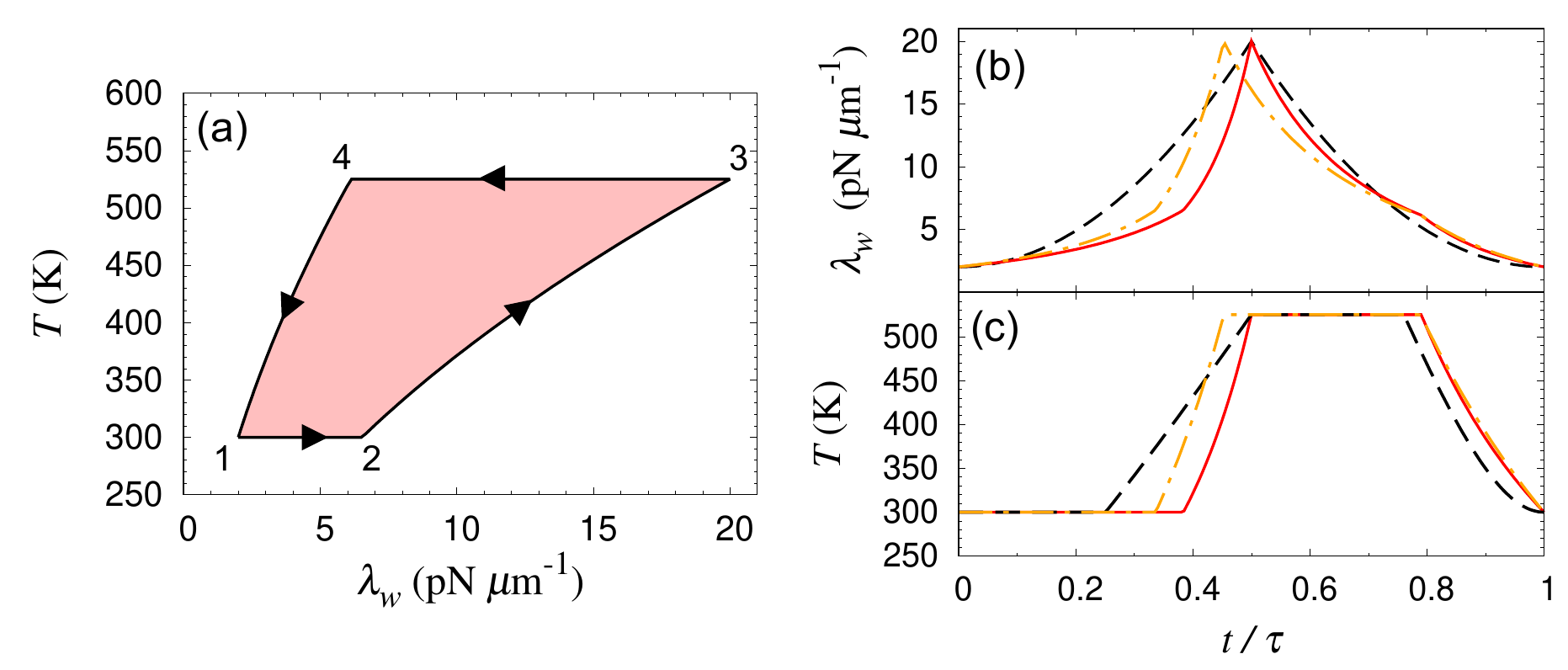}
\caption{Path and protocols for the Brownian Carnot cycle. (a) The path on the $T$-$\lambda_w$ plane in the experiment of Ref.~\cite{Martinez16}. The hot and cold temperatures of the environment in the isothermal strokes are $T_h=525$\,K and $T_c=300$\,K, respectively. The minimum and the maximum values of $\lambda_w$ are $2.0$\,pN\,$\mu$m$^{-1}$ (point 1) and $20.0$\,pN\,$\mu$m$^{-1}$ (point 3), respectively. Right panels: Protocols of (b) $\lambda_w(t)$ and (c) $T(t)$ for one cycle. The red solid lines represent the protocol optimizing $\avg{A}$ (protocol 1), the orange dot-dashed lines represent the one optimizing $\avg{\Delta A^2}$ (protocol 2), and the black dashed lines represent the protocol in the experiment~\cite{Martinez16}.
}
\label{fig:path_protocol}
\end{figure*}

\subsection{Application to the Brownian Carnot engine}\label{sec:brownian_carnot}

As an application of the formalism developed in this work, we consider the experimental realization of a Brownian Carnot engine reported by Mart\'inez {\it et al.}~\cite{Martinez16}. The Brownian Carnot engine is a microscopic heat engine in which a Brownian particle trapped in an optical potential serves as the working substance. Its cycle consists of two isothermal strokes, performed under thermal environment at the hot and cold temperatures $T_h$ and $T_c$, respectively, and two isentropic strokes. Here, the isentropic strokes are defined as processes during which the equilibrium entropy $\avg{S}_{\mathrm{eq}}$ remains constant~\cite{Martinez15,Martinez16} (see also Refs.~\cite{Frim22,Watanabe22}).

Following the experiment, we employ the harmonic potential of Eq.~(\ref{eq:vho}), using its strength $\lambda_w$ as a control parameter. Since the equilibrium entropy of an underdamped Brownian particle in a harmonic potential is given by
\begin{equation}
  \avg{S}_{\mathrm{eq}} = k_{\mathrm{B}} \left[ 1 + \ln{\left(2\pi k_{\mathrm{B}}T \sqrt{\frac{M}{\lambda_w}}\right)}\right]\,,
\end{equation}
the isentropic strokes satisfy
\begin{equation}
  \frac{T^2}{\lambda_w} = \mbox{const.}\label{eq:isentrope}
\end{equation}
The cycle path on the $T$-$\lambda_w$ plane is shown in Fig.~\ref{fig:path_protocol}(a). Each stroke proceeds as follows:
(i) an isothermal compression $(1 \rightarrow 2)$ at $T_c=300$\,K,\,
(ii) an isentropic compression $(2 \rightarrow 3)$,\,
(iii) an isothermal expansion $(3 \rightarrow 4)$ at $T_h=525$\,K,\, and
(iv) an isentropic expansion $(4 \rightarrow 1)$.
The values of $\lambda_w$ at points 1 and 3 are $\lambda_{w, \mathrm{min}} = 2.0$\,pN\,$\mu$m$^{-1}$ and $\lambda_{w, \mathrm{max}} = 20.0$\,pN\,$\mu$m$^{-1}$, respectively~\cite{Martinez16}. The mass of the Brownian particle and the friction coefficient are $M = 5.2 \times 10^{-4}$\,pN\,$\mu$m$^{-1}$\,ms$^{2}$ and $\gamma = 8.4$\,pN\,$\mu$m$^{-1}$\,ms~\cite{Martinez16}.

Along this closed path, whose length is $\mathcal{L}^{(i)}= \mathcal{L}_{C}^{(i)}$, the control parameters $\lambda_w(t)$ and $T(t)$ are driven according to prescribed protocols. In this work, we consider three protocols: a protocol minimizing $\avg{A}$ (protocol 1); a protocol minimizing $\avg{\Delta A^2}$ (protocol 2); and the protocol employed in the experiment of Ref.~\cite{Martinez16}. Protocols 1 and 2 are designed to saturate the geometric bounds given in Eqs.~(\ref{eq:bound1}) and (\ref{eq:bound2}), respectively. More specifically, from the equality condition of the Cauchy-Schwarz inequality, these geometric bounds of $\avg{A}$ and $\avg{\Delta A^2}$ are saturated when the integrand in the definition (\ref{eq:tdlengthi}) of the corresponding thermodynamic length $\mathcal{L}^{(i)}$ remains constant throughout the cycle:
\begin{equation}
  g^{(i)}_{\mu\nu}(t) \dot{\lambda}_\mu \dot{\lambda}_\nu = \mbox{const}\,,\label{eq:sat_cond}
\end{equation} 
where $i=1,\, 2$ correspond to protocols 1 and 2, respectively. For a given cycle duration $\tau$, the time schedules of $\lambda_w(t)$ and $T(t)$ in protocols 1 and 2 are determined by numerically solving
\begin{equation}
  \int_0^\tau dt\, \sqrt{g^{(i)}_{\mu\nu}(t) \dot{\lambda}_\mu(t)\, \dot{\lambda}_\nu(t)} = \mathcal{L}_C^{(i)}\,,
\end{equation}
subject to the constraint (\ref{eq:sat_cond}). The resulting protocols are shown by the red solid lines (protocol 1) and the orange dot-dashed lines (protocol 2) in Figs.~\ref{fig:path_protocol}(b) and \ref{fig:path_protocol}(c). For the experimental protocol, $\lambda_w(t)$ is swept quadratically in time, while $T$ in the isentropic strokes is determined so as to satisfy Eq.~(\ref{eq:isentrope}). This protocol is shown by the black dashed lines in Figs.~\ref{fig:path_protocol}(b) and \ref{fig:path_protocol}(c). Finally, we remark that, in all three protocols, the control parameters $\lambda_w(t)$ and $T(t)$ depend only on the scaled time $t/\tau$ and their functional forms are independent of $\tau$.

\begin{figure}[t!]
\centering
\includegraphics[width=0.99 \columnwidth]{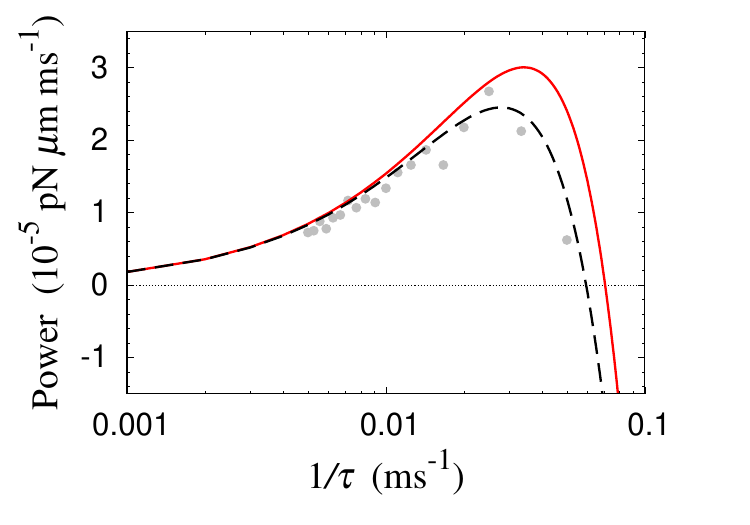}
\caption{Output power $\avg{W}/\tau$ of the Brownian Carnot cycle as a function of the inverse of the cycle period. The lines show theoretical results obtained by numerically solving the full Fokker-Planck equation (\ref{eq:fp_udb}): The red solid line corresponds to protocol~1, and the black dashed line corresponds to the experimental protocol. The gray dots represent experimental data taken from Ref.~\cite{Martinez16}.
}
\label{fig:power}
\end{figure}

The Fokker-Planck equation corresponding to the Langevin equation (\ref{eq:langevin}) is
\begin{equation}
  \frac{\partial}{\partial t} \rho + \frac{p}{M} \frac{\partial}{\partial x} \rho - \lambda_w x \frac{\partial}{\partial p} \rho - \frac{\gamma}{M} \frac{\partial}{\partial p}(p\rho) - \frac{\gamma}{\beta} \frac{\partial^2}{\partial p^2} \rho = 0\,.\label{eq:fp_udb}
\end{equation}
As detailed in Appendix~\ref{app:eoms}, we numerically solve the full Fokker-Planck equation for prescribed protocols of $\lambda_w(t)$ and $T(t)$ discussed above. We first focus on the output power, defined as the average work per cycle divided by the cycle period, $\avg{W}/\tau$. Figure~\ref{fig:power} compares the power obtained with protocol 1 (red solid line) and the experimental protocol (black dashed line). The experimental data (gray dots) are consistent with the theoretical result (black dashed line) obtained by numerically solving the Fokker-Planck equation using the experimental protocol. Since the experimental parameters are used directly in the numerical calculation without any fitting parameters, this agreement supports the validity of the Fokker-Planck description for the present system. Importantly, protocol 1, which minimizes the average dissipation $\avg{A}$, yields a higher output power than the experimental protocol for the same cycle period $\tau$, highlighting the advantage of optimized driving obtained by our formalism.

\begin{figure}[t!]
\centering
\includegraphics[width=0.99 \columnwidth]{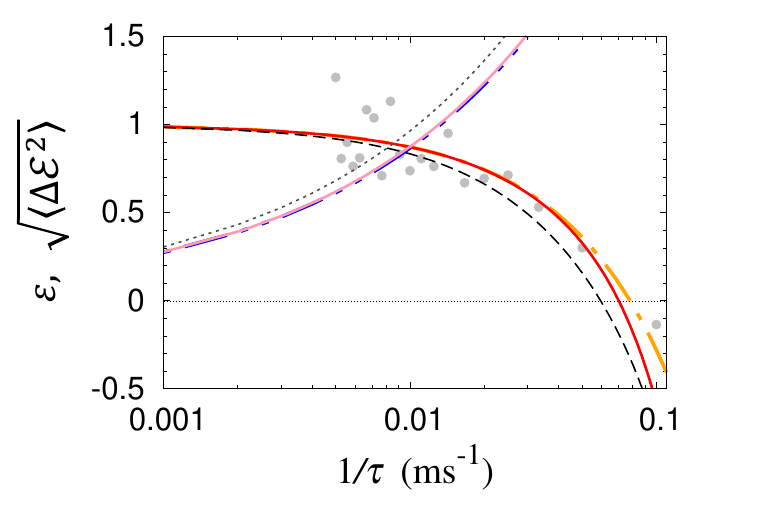}
\caption{Efficiency and its fluctuation of the Brownian Carnot cycle as functions of the inverse cycle period. The three decreasing curves with $1/\tau$ show the efficiency $\epsilon$: The red solid line corresponds to protocol 1, the orange dot-dashed line corresponds to the geometric bound $\epsilon_{\mathrm{geo}}$, and the black dashed line corresponds to the experimental protocol. The three increasing curves with $1/\tau$ show the fluctuation $\sqrt{\avg{\Delta\mathcal{E}^2}}$ of the stochastic efficiency: The pink solid line corresponds to protocol 1, the blue double-dot-dashed line corresponds to the geometric bound $\sqrt{\avg{\Delta\mathcal{E}^2}^{\mathrm{geo}}}$, and the black dotted line corresponds to the experimental protocol. The gray dots represent experimental data of $\epsilon$ taken from Ref.~\cite{Martinez16}.
}
\label{fig:efficiency}
\end{figure}

Next, we study the efficiency and its fluctuation, and compare them with their geometric bounds. We evaluate the efficiency directly from Eq.~(\ref{eq:efficiency}), $\epsilon \equiv \avg{W}/\avg{U}$, by solving the full Fokker-Planck equation without invoking the linear-response approximation, as detailed in Appendix~\ref{app:eoms}. The resulting efficiencies for protocol 1 and for the experimental protocol are shown by the red solid line and the black dashed line, respectively, in Fig.~\ref{fig:efficiency}. The efficiency achieved by protocol 1 is always higher than that obtained with the experimental protocol for any value of $\tau$, demonstrating that the performance is indeed improved by employing protocol 1 along the same path and with the same $\tau$. The geometric bound of the efficiency, $\epsilon_{\mathrm{geo}}$ [Eq.~(\ref{eq:geobound1})], derived within the linear-response approximation, is shown by the orange dot-dashed line. The deviation of the red solid line from the geometric bound for $1/\tau \gtrsim 0.05$ms$^{-1}$ indicates that this regime lies beyond the linear-response regime, whereas the good agreement between the two curves for $1/\tau \lesssim 0.05$ms$^{-1}$ confirms the validity of the linear-response approximation in this region. The validity of the linear-response approximation can be assessed by comparing the driving timescale with the effective correlation times of the system. When the former becomes comparable to or shorter than the latter, the slow-driving condition is violated, leading to the breakdown of the linear-response approximation. For the underdamped Brownian particle considered here, the relevant effective correlation times are given by Eqs.~(\ref{eq:tau_ww})--(\ref{eq:tau_uu}). Along the present cycle path, the largest value among these correlation times is approximately $2.1$\,ms. On the other hand, for the cycle period $\tau \simeq 20$\,ms corresponding to $1/\tau \simeq 0.05$\,ms$^{-1}$, the shortest stroke duration for protocol 1 is approximately $2.3$\,ms. Therefore, the driving timescale becomes comparable to the largest effective correlation time, indicating the breakdown of the slow-driving condition. This estimate is consistent with the onset of the deviation from the geometric bound discussed above. For longer cycle periods, the driving timescale increases while the relaxation times remain unchanged, thereby restoring the validity of the linear-response approximation.

We next turn to the fluctuation of the stochastic efficiency, characterized by $\sqrt{\avg{\Delta \mathcal{E}^2}}$. This quantity is evaluated from Eq.~(\ref{eq:var_efficiency}) together with $\avg{\Delta A^2}$ given by Eqs.~(\ref{eq:vara}) and (\ref{eq:g2munu}) within the linear-response approximation. 
The resulting $\sqrt{\avg{\Delta \mathcal{E}^2}}$ for protocol 1 and that for the experimental protocol are shown by the pink solid line and the black dotted line, respectively, in Fig.~\ref{fig:efficiency}. We find that protocol 1 yields systematically smaller values of $\sqrt{\avg{\Delta \mathcal{E}^2}}$ than the experimental protocol, indicating that protocol 1 outperforms the experimental one also in terms of the fluctuation of the efficiency. It is also noted that $\sqrt{\avg{\Delta \mathcal{E}^2}}$ is comparable to, or even larger than, unity in the region of $1/\tau$ explored in the experiment. Correspondingly, the experimental data of the efficiency (gray dots) exhibit substantial scatter around the theoretical prediction shown by the black dashed line~\cite{note:sample}. The geometric bound of the fluctuation of the efficiency, $\avg{\Delta \mathcal{E}^2}^{\mathrm{geo}}$ [Eq.~(\ref{eq:geobound2})], is shown by the blue double-dot-dashed line. Although this bound is strictly saturated only by protocol 2, which differs from protocol 1 as shown in Figs.~\ref{fig:path_protocol}(b) and \ref{fig:path_protocol}(c), the fluctuation $\sqrt{\avg{\Delta \mathcal{E}^2}}$ for protocol 1 is already very close to the geometric bound in the present setting.

\section{Conclusion and prospects}\label{sec:conclusion}

In this work, we have developed a geometric framework for the finite-time thermodynamics of microscopic cyclic heat engines in the presence of fluctuations. Our formulation provides a unified description of dissipation in terms of equilibrium correlation functions and enables a geometric characterization of both the mean dissipated availability and its variance.

Within the linear-response regime and under a time-local approximation, we have shown that the average dissipation and its fluctuations can be expressed in terms of metric tensors constructed from equilibrium correlation functions. In particular, the relevant correlation times entering the time-local description are characterized by effective correlation times defined in Eq.~(\ref{eq:taubar}). These metrics, given explicitly in Eqs.~(\ref{eq:g1munu_final}) and (\ref{eq:g1g2rel}), define thermodynamic lengths in the space of control parameters, which in turn determine geometric bounds on the mean dissipation and its variance during a finite-time cycle. Notably, the fluctuation metric $g^{(2)}_{\mu\nu}$ is universally proportional to the dissipation metric $g^{(1)}_{\mu\nu}$, revealing a direct geometric relation between dissipation and stochastic fluctuations. The present formalism applies to a wide class of systems, including Markov jump processes as well as overdamped and underdamped Brownian dynamics with multiple relaxation timescales.

We have illustrated the general formalism through several representative examples, including a classical two-level system (Sec.~\ref{sec:markovjump}), an overdamped Brownian particle in a power-law potential (Sec.~\ref{sec:overdampedbp}), and an underdamped Brownian Carnot engine in a harmonic potential (Sec.~\ref{sec:underdampedbp}). These examples demonstrate how the dynamical properties of the system are encoded in the geometric metrics and how the resulting framework provides a unified description of dissipation and its fluctuations in finite-time cyclic processes.

The present results suggest that the performance of microscopic heat engines can be constrained and understood from a geometric perspective that simultaneously incorporates dissipation and fluctuations. In particular, the geometric bounds derived in this work imply that the thermodynamic length of the cycle path in control-parameter space plays a central role in determining not only the mean dissipation but also the magnitude of stochastic fluctuations [Eqs.~(\ref{eq:bound1}) and (\ref{eq:bound2})].

Combining the geometric bounds for the mean dissipation and its variance yields a lower bound on the relative fluctuation of stochastic efficiency [Eq.~(\ref{eq:geobound_flucteff})], demonstrating that efficiency fluctuations cannot be made arbitrarily small for a finite cycle duration. This bound is conceptually distinct from conventional trade-off relations such as thermodynamic uncertainty relations, as it arises from geometric properties of the control protocol rather than solely from entropy production.

More broadly, the present work suggests that finite-time thermodynamics may admit a systematic geometric structure beyond average quantities. In this view, thermodynamic metrics derived from equilibrium correlation functions provide a bridge between microscopic dynamics and macroscopic performance constraints of small-scale thermal machines.

The present framework opens several promising directions for future research. A natural extension concerns higher-order fluctuations of dissipation. While the present work focuses on the variance of the dissipated availability, higher moments involve multitime correlation functions of thermodynamic forces.

Within the time-local approximation used in this work, two-time correlations effectively reduce to delta-function correlations reflecting short-time relaxation dynamics. Extending this idea to higher-order correlations suggests that multitime correlation functions may be represented as combinations of delta functions in the time-local limit. Beyond the time-local limit, such correlations are expected to admit systematic representations in terms of relaxation modes associated with the dynamical spectrum of the system. Developing this structure may provide a route toward a geometric description of higher-order dissipation statistics and non-Gaussian fluctuations in microscopic heat engines.

Ultimately, this perspective raises the possibility that finite-time thermodynamics of small systems may be organized by a hierarchy of geometric structures derived from multitime correlation functions, providing a unified framework for understanding dissipation, fluctuations, and control in microscopic thermal machines that can be realized in stochastic thermodynamic experiments.

\begin{acknowledgments} 
This work was supported by NSF of China (Grants No.~12375039 and No.~11975199), by the Zhejiang Provincial Natural Science Foundation Key Project (Grant No.~LZ19A050001), and by the Zhejiang University 100 Plan.
\end{acknowledgments}

\section*{Data Availability}
The data and numerical codes supporting the findings of this study are available from the authors upon reasonable request.

\appendix

\section{Derivation of $\avg{\Delta A^2}$ in the time-local approximation}\label{app:derivation}

In this appendix, we provide a detailed derivation of Eqs.~(\ref{eq:vara}) and (\ref{eq:g2munu}). The derivation closely follows Ref.~\cite{Watanabe22}, but is reproduced here for completeness and to clarify the role of the time-local approximation and boundary contributions.

In the decomposition of the variance $\avg{\Delta A^2}$ of $A$, $\avg{\Delta A^2} = \avg{\Delta U^2} + \avg{\Delta W^2} - 2 \avg{\Delta U\, \Delta W}$, we first evaluate the contribution from the work. Using Eq.~(\ref{eq:corrgen}) together with the closed-cycle condition that the phase-space distribution function satisfies $\mathcal{P}(\Gamma, 0)=\mathcal{P}(\Gamma, \tau)$, we obtain
\begin{align}
  \avg{\Delta W^2} =& \int_0^\tau dt \int_0^\tau dt'\, \left[ \avg{X_w(t) X_w(t')} - \avg{X_w(t)} \avg{X_w(t')} \right]\,\nonumber\\
  & \times \dot{\lambda}_w(t)\, \dot{\lambda}_w(t')\nonumber\\
  =&\, 2 \int_0^\tau dt\, \avg{\Delta X_w^2(t)}_{\mathrm{eq}}\, \overline{\tau}_{ww}(t)\, \dot{\lambda}_w^2(t)\,.\label{eq:varw}
\end{align}

Next, we consider the contribution associated with $U$. By integrating by parts, the quantity $U$ defined in Eq.~(\ref{eq:u}) can be rewritten as
\begin{equation}
  U = -\int_0^\tau X_u(t)\, \dot{\lambda}_u(t)\, dt + [X_u(\tau)-X_u(0)]\, \lambda_u(0)\,,
\end{equation}
where we have used $\lambda_u(\tau) = \lambda_u(0)$, while $X_u(\tau) \ne X_u(0)$ in general. Using again Eq.~(\ref{eq:corrgen}) and the closed-cycle condition, we obtain
\begin{align}
  \avg{\Delta U^2} =\, 2 \int_0^\tau dt\, \avg{\Delta X_u^2(t)}_{\mathrm{eq}}\, \overline{\tau}_{uu}(t)\, \dot{\lambda}_u^2(t) + 2\, \avg{\Delta X_u^2(0)}_{\mathrm{eq}}\, \lambda_u^2(0)\,.\label{eq:varu_withend}
\end{align}
The second term originates from the end points of the cycle. Since it vanishes upon averaging over many cycles, we neglect it in the following and focus on the bulk contribution:
\begin{align}
  \avg{\Delta U^2} \simeq 2 \int_0^\tau dt\, \avg{\Delta X_u^2(t)}_{\mathrm{eq}}\, \overline{\tau}_{uu}(t)\, \dot{\lambda}_u^2(t)\,.\label{eq:varu}
\end{align}
Similarly, the covariance between $W$ and $U$ is given by
\begin{align}
  \avg{\Delta W \Delta U} =&\, -2 \int_0^\tau dt\, \avg{\Delta X_w(t)\, \Delta X_u(t)}_{\mathrm{eq}}\, \overline{\tau}_{wu}(t)\, \dot{\lambda}_w(t)\, \dot{\lambda}_u(t)\,.\label{eq:covwu}
\end{align}

Combining Eqs.~(\ref{eq:varw}), (\ref{eq:varu}), and (\ref{eq:covwu}), we obtain the variance of $A$ in the time-local approximation,
\begin{align}
  \avg{\Delta A^2} = \int_0^\tau dt\, g^{(2)}_{\mu\nu}(t)\, \dot{\lambda}_\mu(t)\, \dot{\lambda}_\nu(t)\,,
\end{align}
with
\begin{align}
  g^{(2)}_{\mu\nu}(t) \equiv 2\, \overline{\tau}_{\mu\nu}(t)\, \avg{\Delta X_\mu(t) \Delta X_\nu(t)}_{\mathrm{eq}} = 2\, \overline{\tau}_{\mu\nu}(t)\, \sigma_{\mu\nu}(t)\,.
\end{align}

\section{Spectral decomposition and correlation functions for the classical two-level system}\label{app:corrfunc}

For the transition-rate matrix $R$ given by Eq.~(\ref{eq:ratematrix_tls}),
\begin{equation}
  R =
  \begin{pmatrix}
    -r_- & r_+ \\
    r_- & -r_+
  \end{pmatrix}\,,\label{appeq:ratematrix_tls}
\end{equation}
the steady-state distribution is [Eq.~(\ref{eq:pi_pm})]:
\begin{equation}
  \pi_\pm = \frac{r_\pm}{r_+ + r_-} = \frac{e^{\mp\beta\Delta/2}}{2 \cosh{\left(\dfrac{\beta\Delta}{2}\right)}}\,.\label{appeq:pi_pm}
\end{equation}
The eigenvalues of $R$ are as follows [Eqs.~(\ref{eq:lambda0_tls}) and (\ref{eq:lambda1_tls})]: $\Lambda_0 = 0$ and $\Lambda_1 = r_+ + r_- = 2\gamma \cosh{(\beta\Delta/2)}$. Their corresponding right eigenvectors, $R\phi^{(0)} = 0$ and $R\phi^{(1)}=-\Lambda_1 \phi^{(1)}$, are
\begin{equation}
  \phi^{(0)} =
  \begin{pmatrix}
    \pi_+ \\
    \pi_-
  \end{pmatrix}
  \quad\mbox{and}\quad
  \phi^{(1)} =
  \begin{pmatrix}
    1 \\
    -1
  \end{pmatrix}\,,\label{appeq:phik}
\end{equation}
and left eigenvectors, $(\psi^{(0)})^{\mathsf{T}} R = 0$ and $(\psi^{(1)})^{\mathsf{T}} R = -\Lambda_1 (\psi^{(1)})^{\mathsf{T}}$, are 
\begin{equation}
  \psi^{(0)} =
  \begin{pmatrix}
    1\\
    1
  \end{pmatrix}
  \quad\mbox{and}\quad
  \psi^{(1)} =
  \begin{pmatrix}
    \pi_-\\
    -\pi_+
  \end{pmatrix}\,.\label{appeq:psik}
\end{equation}
We can readily confirm that these eigenvectors indeed satisfy the biorthonormality condition $(\psi^{(k)})^{\mathsf{T}} \phi^{(l)} = \delta_{k, l}$ [Eq.~(\ref{eq:biorthonormal})] and the completeness relation $\sum_{k=0,1} \phi^{(k)} (\psi^{(k)})^{\mathsf{T}} = I$ [Eq.~(\ref{eq:complete})].

For the eigenvectors $\phi^{(k)}$ and $\psi^{(k)}$ given by Eqs.~(\ref{appeq:phik}) and (\ref{appeq:psik}), terms for each $k$ in the two-time correlation function (\ref{eq:corrfunc_ab}) reduce to
\begin{align}
  \sum_{n=\pm} A_n \phi_n^{(0)} &= \sum_{n=\pm} A_n \pi_n = \avg{A}_{\mathrm{eq}}\,,\label{appeq:anphin0}\\
  \sum_{m=\pm} B_m \psi_m^{(0)} &= \sum_{m=\pm} B_m \pi_m = \avg{B}_{\mathrm{eq}}\,,\label{appeq:bmpsim0}
\end{align}
and
\begin{align}
  \sum_{n=\pm} A_n \phi_n^{(1)} &= A_+ - A_-\,,\label{appeq:anphin1}\\
  \sum_{m=\pm} B_m \psi_m^{(1)} &= (B_+ - B_-) \pi_+ \pi_-\,.\label{appeq:bmpsim1}
\end{align}
For $A=B=\sigma = (+1, -1)^{\mathsf{T}}$, using Eqs.~(\ref{appeq:anphin0})--(\ref{appeq:bmpsim1}) and $\pi_+ \pi_- = (1-m^2)/4$ with $m\equiv \avg{\sigma}_{\mathrm{eq}} = \pi_+ - \pi_-$ being the equilibrium magnetization [Eq.~(\ref{eq:m})], the two-time correlation function (\ref{eq:corrfunc_ab}) reduces to
\begin{align}
  \avg{\sigma(t)\, \sigma(0)}_{\mathrm{eq}} = m^2 + (1-m^2) e^{-\Lambda_1 t}\,.\label{appeq:corrfunc_sigma}
\end{align}

From this result, we next evaluate the equilibrium two-time correlation functions of the generalized forces $X_w = -\partial H(\sigma)/\partial \Delta = -\sigma/2$ and $X_u = S(\sigma) = -k_{\mathrm{B}} \ln{P_\sigma}$. For $X_w$, we obtain
\begin{align}
  \avg{\Delta X_w(t)\, \Delta X_w(0)}_{\mathrm{eq}} &= \frac{1}{4} \left[ \avg{\sigma(t)\, \sigma(0)}_{\mathrm{eq}} - \avg{\sigma}_{\mathrm{eq}}^2 \right]\nonumber\\
  &= \frac{1}{4} (1-m^2) e^{-\Lambda_1 t}\,,\label{appeq:xwxw_tls}
\end{align}
where we have used Eq.~(\ref{appeq:corrfunc_sigma}) from the first and the second line.

For the steady state $P_\sigma = \pi_\sigma$ given by Eq.~(\ref{appeq:pi_pm}), the stochastic entropy $X_u = S(\sigma)$ reads
\begin{equation}
  S(\sigma) = -k_{\mathrm{B}} \ln{\pi_\sigma} = k_{\mathrm{B}} \ln{\left(2 \cosh{\frac{\beta\Delta}{2}}\right)} + k_{\mathrm{B}} \frac{\beta\Delta}{2} \sigma\,.
\end{equation}
Thus, the two-time correlation function of $X_u$ can be written as
\begin{align}
  \avg{\Delta X_u(t)\, \Delta X_u(0)}_{\mathrm{eq}} &= \avg{S(t)\, S(0)}_{\mathrm{eq}} - \avg{S}_{\mathrm{eq}}^2\nonumber\\
  &= k_{\mathrm{B}}^2 \frac{\beta^2\Delta^2}{4} (1-m^2) e^{-\Lambda_1 t}\,,\label{appeq:xuxu_tls}
\end{align}
and the cross-correlation function between $X_w$ and $X_u$ as
\begin{align}
  \avg{\Delta X_w(t)\, \Delta X_u(0)}_{\mathrm{eq}} &= \avg{\Delta X_u(t)\, \Delta
    X_w(0)}_{\mathrm{eq}} \nonumber\\
  &= -k_{\mathrm{B}} \frac{\beta\Delta}{4} \left[\avg{\sigma(t)\, \sigma(0)}_{\mathrm{eq}} - \avg{\sigma}_{\mathrm{eq}}^2\right]\nonumber\\
  &= -k_{\mathrm{B}} \frac{\beta\Delta}{4} (1-m^2) e^{-\Lambda_1 t}\,.\label{appeq:xwxu_tls}
\end{align}

From the resulting correlation functions $\avg{\Delta X_\mu(t)\, \Delta X_\nu(0)}_{\mathrm{eq}}$ given by Eqs.~(\ref{appeq:xwxw_tls}), (\ref{appeq:xuxu_tls}), and (\ref{appeq:xwxu_tls}), we obtain
\begin{align}
  \tau_{ww} &= \tau_{uu} = \tau_{wu} = \tau_{uw} = 1/\Lambda_1\,,\\
  C_{ww} &= C_{uu} = C_{wu} = C_{uw} = 1\,,
\end{align}
and the covariance tensor $\sigma_{\mu\nu}$ as
\begin{align}
  \sigma_{ww} &= \frac{1}{4} (1-m^2)\,,\\
  \sigma_{uu} &= k_{\mathrm{B}}^2 \frac{\beta^2\Delta^2}{4} (1-m^2)\,,\\
  \sigma_{wu} &= \sigma_{uw} = -k_{\mathrm{B}}\frac{\beta\Delta}{4} (1-m^2)\,.
\end{align}

\section{Spectral decomposition of the Fokker-Planck operator}\label{app:decomp_fpop}

In this section, we provide details of the spectral decomposition of the Fokker-Planck operator and the eigenfunction expansion of the transition probability~\cite{Risken_book}.
For the overdamped Brownian particle in a 1D potential $V_{\lambda_w}(x)$, the Fokker-Planck operator $L_{\mathrm{FP}}$ is defined as
\begin{align}
  L_{\rm FP} \equiv -\frac{\partial}{\partial x} D^{(1)}(x) + \frac{\partial^2}{\partial x^2} D^{(2)}(x)\,,
\end{align}
with the drift coefficient
\begin{align}
  D^{(1)} \equiv \frac{F(x)}{\gamma} = -\frac{1}{\gamma} \frac{\partial V_{\lambda_w}}{\partial x}\,
\end{align}
and the diffusion coefficient
\begin{align}
  D^{(2)} \equiv \frac{k_{\mathrm{B}}T}{\gamma}\,.
\end{align}
The operator $L_{\mathrm{FP}}$ can be transformed into a Hermitian operator $L_{\mathrm{H}}$ by a similarity transformation:
\begin{align}
  L_{\mathrm{H}} = e^{\Phi/2}\, L_{\mathrm{FP}}\, e^{-\Phi/2}\,,
\end{align}
where
\begin{equation}
  \Phi \equiv \ln{D^{(2)}} - \int^x \frac{D^{(1)}(x')}{D^{(2)}(x')} dx'
  = \ln{\frac{k_{\mathrm{B}}T}{\gamma}} + \frac{V_{\lambda_w}(x)}{k_{\mathrm{B}}T}\,,
\end{equation}
so that
\begin{align}
  e^\Phi = \frac{k_{\mathrm{B}}T}{\gamma}\, \exp{\left[\frac{V_{\lambda_w}(x)}{k_{\mathrm{B}}T}\right]}\,.
\end{align}
With $\Phi$, the Fokker-Planck operator can be written as
\begin{align}
  L_{\mathrm{FP}}\, \cdots = \frac{\partial}{\partial x} D^{(2)} e^{-\Phi(x)} \frac{\partial}{\partial x} e^{\Phi(x)}\cdots\,,
\end{align}
and thus the transformed Hermitian operator $L_{\mathrm{H}}$ reads
\begin{align}
  L_{\mathrm{H}}\, \cdots &= e^{\Phi/2} L_{\mathrm{FP}} e^{-\Phi/2} \cdots\nonumber\\
  &= e^{\Phi(x)/2} \frac{\partial}{\partial x} D^{(2)} e^{-\Phi(x)} \frac{\partial}{\partial x} e^{\Phi(x)/2}\cdots\,.
\end{align}
The operators $L_{\mathrm{FP}}$ and $L_{\mathrm{H}}$ share the same eigenvalues $\{\Lambda_i\}$. Denoting the $i$th eigenfunction of $L_{\mathrm{FP}}$ and $L_{\mathrm{H}}$ by $\phi_i$ and $\tilde{\phi}_i$, respectively, we thus have
\begin{align}
  L_{\rm FP} \phi_i &= -\Lambda_i \phi_i\,,\label{eq:eigeneqLfp}\\
  L_{\rm H} \tilde{\phi}_i &= -\Lambda_i \tilde{\phi}_i\,,\label{eq:eigeneqLh}
\end{align}
with
\begin{align}
  \tilde{\phi}_i = e^{\Phi/2} \phi_i\,.\label{eq:psiphi}
\end{align}
Here, all eigenvalues are non-negative, $\Lambda_i \ge 0$, and are sorted in ascending order: $0 = \Lambda_0 < \Lambda_1 \le \Lambda_2 \le \cdots$.

The stationary solution of the Fokker-Planck equation is $\rho_{\mathrm{eq}}(x) \propto e^{-\beta V_{\lambda_w}} \propto e^{-\Phi}$, which corresponds to the zero eigenvalue eigenstate $\phi_0$: $\rho_{\mathrm{eq}}(x) = N_0 e^{-\Phi} = \sqrt{N_0} \phi_0$ and $\phi_0 = \sqrt{N_0} e^{-\Phi}$, where $N_0$ is the normalization constant. Using Eq.~(\ref{eq:psiphi}), we then have
\begin{align}
  \tilde{\phi}_0 = \sqrt{N_0}\, e^{-\Phi/2}\,,\label{eq:psi0}
\end{align}
and thus
\begin{align}
  \rho_{\mathrm{eq}}(x) = \tilde{\phi}_0^2(x)\,.\label{eq:peq}
\end{align}

Since eigenfunctions of Hermitian operators form a complete set, we can decompose the correlation function in terms of the eigenfunctions $\{\tilde{\phi}_i\}$ of $L_{\mathrm{H}}$. Using the completeness of the eigenbasis set $\{\tilde{\phi}_i\}$,
\begin{align}
  \delta(x-x') &= \sum_{i=0}^\infty \tilde{\phi}_i(x)\, \tilde{\phi}_i(x')\nonumber\\
  &= \exp[\Phi(x')]\, \sum_{i=0}^\infty \phi_i(x)\, \phi_i(x'),
\end{align}
the transition probability $\rho(x,t|x',t')$ for $t>t'$ can be written as
\begin{align}
  \rho(x,t|x',t') &= e^{(t-t')\, L_{\mathrm{FP}}(x)} \delta(x-x')\nonumber\\
  & = e^{\Phi(x')} e^{(t-t')\, L_{\mathrm{FP}}(x)} \sum_{i=0}^\infty \phi_i(x)\, \phi_i(x')\nonumber\\
  & = e^{\Phi(x')/2} e^{-\Phi(x)/2} \sum_{i=0}^\infty e^{-\Lambda_i\, (t-t')} \tilde{\phi}_i(x)\, \tilde{\phi}_i(x')\nonumber\\
  & = \frac{\tilde{\phi}_0(x)}{\tilde{\phi}_0(x')} \sum_{i=0}^\infty e^{-\Lambda_i\, (t-t')} \tilde{\phi}_i(x)\, \tilde{\phi}_i(x')\,.
\end{align}
Here, we have obtained Eq.~(\ref{eq:transition_prob}), which gives the eigenfunction expansion of the transition probability.

\section{Sturm-Liouville eigenvalue problem}\label{app:slproblem}

In the Sturm-Liouville eigenvalue problem, we seek solutions to real second-order linear ordinary differential equations of the following form (see, e.g., Ref.~\cite{Risken_book} for details):
\begin{align}
  \frac{d}{dx}\left[u(x)\frac{dy_i(x)}{dx}\right] + v(x)\, y_i(x) = - \Lambda_i\, w(x)\, y_i(x)\,,\label{eq:sturm-liouville}
\end{align}
which is known as the Sturm-Liouville equation.
Here, $u(x)$, $v(x)$, and $w(x)$ are given coefficient functions, continuous on the interval $x \in [a,b]$, with $u(x)>0$ and $w(x)>0$; $y_i(x)$ is the eigenfunction corresponding to the eigenvalue $\Lambda_i$. The coefficient function $w$ is referred to as the weight function, and the normalized eigenfunctions $\{y_i\}$ form an orthonormal basis set with respect to the $w$-weighted inner product,
\begin{align}
  \int_a^b y_i(x)\, y_j(x)\, w(x)\, dx = \delta_{ij}\,.
\end{align}

Regarding the specific problem discussed in Sec.~\ref{sec:overdampedbp}, the eigenvalue equation (\ref{eq:eigeneqLh}) of $L_{\rm H}$ [and thus Eq.~(\ref{eq:eigeneqLfp}) for $L_{\rm FP}$ as well] can be identified as the Sturm-Liouville equation for $y_i = e^{\Phi/2} \tilde{\phi}_i$ (or, equivalently, $y_i= e^{\Phi} \phi_i$) with
\begin{align}
  u(x) &= \frac{k_{\mathrm{B}}T}{\gamma} e^{-\Phi} = \exp{\left[-\frac{V_{\lambda_w}(x)}{k_{\mathrm{B}}T}\right]}\,,\label{eq:coeff_u}\\
  v(x) &= 0\,,\label{eq:coeff_v}\\
  w(x) &= e^{-\Phi} = \frac{\gamma}{k_{\mathrm{B}}T} \exp{\left[-\frac{V_{\lambda_w}(x)}{k_{\mathrm{B}}T}\right]}\,,\label{eq:coeff_w}
\end{align}
for $x \in (-\infty,\infty)$ under natural boundary conditions.
Namely, Eq.~(\ref{eq:eigeneqLh}) reduces to the Sturm-Liouville equation (\ref{eq:sturm-liouville}) as
\begin{widetext}
\begin{align}
  L_{\rm H} \left( e^{-\Phi(x)/2} y_i(x) \right) = e^{\Phi(x)/2} \frac{d}{dx}\left[ \frac{k_{\mathrm{B}}T}{\gamma} e^{-\Phi(x)} \frac{d}{dx} y_i(x) \right] = -\Lambda_i \left(e^{-\Phi(x)/2} y_i(x) \right)\,.
\end{align}
\end{widetext}
Accordingly, the eigenfunction $\tilde{\phi}_i(x)$ of the Hermitian operator is expressed as
\begin{align}
  \tilde{\phi}_i(x) = e^{-\Phi(x)/2} y_i(x) = \sqrt{\frac{\gamma}{k_{\mathrm{B}}T}}\, \exp{\left[-\frac{V_{\lambda_w}(x)}{2k_{\mathrm{B}}T}\right]}\, y_i(x)\,.
\end{align}

\renewcommand{\arraystretch}{1.4}
\begin{table}[!b]
  \caption{Numerical values of $\sum_{i\ge1}c_i/\Lambda_i$ for an overdamped Brownian particle in a 1D power-law potential $V_{\lambda_w}$ with exponent $n$. All the other parameters are set to unity: $\lambda_w = \gamma = k_{\mathrm{B}}T = 1$.}
\begin{tabular}{cl|cl}\hline
  $n$ & $\quad\sum_{i\ge1} c_i/\Lambda_i\quad$ & $n$ & $\quad\sum_{i\ge1} c_i/\Lambda_i\quad$ \\ \hline\hline
  1  &\quad 1            &   13 &\quad 0.413\,455\\
  2  &\quad 0.675\,978   &   14 &\quad 0.409\,067\\
  3  &\quad 0.578\,617   &   15 &\quad 0.405\,191\\
  4  &\quad 0.529\,150   &   16 &\quad 0.401\,738\\
  5  &\quad 0.498\,377   &   17 &\quad 0.398\,641\\
  6  &\quad 0.477\,055   &   18 &\quad 0.395\,846\\
  7  &\quad 0.461\,254   &   19 &\quad 0.393\,30 \\
  8  &\quad 0.448\,995   &   20 &\quad 0.390\,99 \\
  9  &\quad 0.439\,160   &   30 &\quad 0.3754  \\
  10 &\quad 0.431\,067   &   40 &\quad 0.3669  \\
  11 &\quad 0.424\,271   &   50 &\quad 0.3615  \\
  12 &\quad 0.418\,471   & & \\ \hline
\end{tabular}\label{tab:tautilden}
\end{table}

The Sturm-Liouville formulation provides a general procedure for determining the effective correlation times $\overline{\tau}_{\mu\nu}$. By solving the Sturm-Liouville equation (\ref{eq:sturm-liouville}) with Eqs.~(\ref{eq:coeff_u})--(\ref{eq:coeff_w}), we obtain the eigenvalues $\{\Lambda_i\}$ and eigenfunctions $\{ y_i\}$. Using the solutions of $y_i$, the expansion coefficients $c_i$ are calculated from Eq.~(\ref{eq:ci}). The dimensionless quantity $\left[\sum_{i\ge 1}c_i/\Lambda_i\right]_{\lambda_w = \gamma = k_{\mathrm{B}}T = 1}$ depends only on the exponent $n$. We define
\begin{equation}
  f_n \equiv \left[\sum_{i=1}^\infty \frac{c_i}{\Lambda_i}\right]_{\lambda_w=\gamma=k_{\mathrm{B}}T=1}\,.\label{eq:sum_taui}
\end{equation}
The numerical values of $f_n$ are listed in Table~\ref{tab:tautilden}. We also obtain
\begin{equation}
  \left[\sum_{i=1}^\infty c_i\right]_{\lambda_w=\gamma=k_{\mathrm{B}}T=1} = 2n\,.\label{eq:sum_ci}
\end{equation}

Using Eqs.~(\ref{eq:sum_taui}) and (\ref{eq:sum_ci}) together with the scaling relation for the eigenvalues, $\Lambda_i \propto \gamma^{-1} (k_{\mathrm{B}}T)^{1-\frac{1}{n}}\lambda_w^{\frac{\alpha}{n}}$ [Eq.~(\ref{eq:scaling})], the effective correlation times $\overline{\tau}_{\mu\nu}$ given by Eq.~(\ref{eq:taubar_overdampedbp}) are finally obtained:
\begin{align}
  \overline{\tau}_{\mu\nu} &= \gamma\, (k_{\mathrm{B}}T)^{-1+\frac{1}{n}} \lambda_w^{-\frac{\alpha}{n}}\, \left[\frac{1}{\sum_{i=1}^\infty c_i} \sum_{i=1}^\infty \frac{c_i}{\Lambda_i}\right]_{\lambda_w=\gamma=k_{\mathrm{B}}T=1}\nonumber\\
  &= \frac{f_n}{2n} \gamma\, (k_{\mathrm{B}}T)^{-1+\frac{1}{n}} \lambda_w^{-\frac{\alpha}{n}}\,.
\end{align}

For the power-law potential considered here, the coefficient $f_n$ can be derived analytically by using the backward generator, as shown in Appendix~\ref{app:fn}. The resulting analytical expression is
\begin{equation}
  f_n = (2n)^{\frac{1}{n}}\,\frac{\Gamma\left(\frac{3}{2n}\right)}{\Gamma\left(\frac{1}{2n}\right)}\,.
\end{equation}
The numerical values listed in Table~\ref{tab:tautilden} are in agreement with this analytical expression (see Fig.~\ref{fig:fn}).

\section{Analytical derivation of $\overline{\tau}_{\mu\nu}$ and $f_n$ by a backward generator}\label{app:fn}

In Appendix~\ref{app:slproblem}, the effective correlation times $\overline{\tau}_{\mu\nu}$ were evaluated by solving the Sturm-Liouville problem. For the overdamped Brownian particle in a 1D power-law potential, they can in fact be evaluated analytically by using the backward generator. This section presents the derivation.

We consider an overdamped Brownian particle in a 1D power-law potential given by Eq.~(\ref{eq:powerlawpot}), $V_{\lambda_w}(x) = \lambda_w^{\alpha} x^{2n}/ (2n)$, whose dynamics is governed by the Fokker-Planck equation~(\ref{eq:FPeq}). To derive an analytical expression for $\overline{\tau}_{\mu\nu}$ and the coefficient $f_n$, it is convenient to employ the backward generator $\mathcal{L}$ associated with the Fokker-Planck operator given by Eq.~(\ref{eq:FPop}):
\begin{equation}
    {\mathcal L}
    =
    \frac{k_{\mathrm B}T}{\gamma}\partial_x^2
    -
    \frac{\lambda_w^\alpha}{\gamma}x^{2n-1}\partial_x\,.
    \label{eq:backward-generator}
\end{equation}

From Eqs.~(\ref{eq:corrmulti}) and (\ref{eq:taubar}), $\overline{\tau}_{\mu\nu}$ can be expressed by the time integral of the two-time correlation function as
\begin{equation}
      \overline{\tau}_{\mu\nu} = \dfrac{\displaystyle \int_0^{\infty}dt\, \avg{\Delta X_\mu(t)\, \Delta X_\nu(0)}_{\mathrm{eq}}}{\avg{\Delta X_\mu(0)\, \Delta X_\nu(0)}_{\mathrm{eq}}}.
      \label{eq:taubar_by_timeintegral}
\end{equation}
We define
\begin{equation}
    h(x) \equiv x^{2n} - \avg{x^{2n}}_{\mathrm{eq}}\,,
\end{equation}
and the relevant time-integrated equilibrium correlation function is
\begin{equation}
    I \equiv  \int_0^\infty dt\, \avg{h(X_t)\, h(X_0)}_{\mathrm{eq}}\,.
    \label{eq:I-def}
\end{equation}
Using the backward generator $\mathcal{L}$, $I$ can formally be written as
\begin{equation}
  I = \int_0^\infty dt\, \avg{h \, e^{\mathcal{L}t}h}_{\mathrm{eq}}\,.
\end{equation}

To evaluate this integral analytically, we seek a function $\varphi(x)$ satisfying the Poisson equation $\mathcal{L}\varphi(x) = -h(x)$. Noting that
\begin{align}
    {\mathcal L} x^2 = \frac{2k_{\mathrm B}T}{\gamma}
    - \frac{2\lambda_w^\alpha}{\gamma}x^{2n}\,,
\end{align}
we can readily see
\begin{align}
    {\mathcal L}
    \left(
        \frac{\gamma}{2\lambda_w^\alpha}x^2
    \right)
    & = 
    \frac{k_{\mathrm B}T}{\lambda_w^\alpha} - x^{2n} \\
    & = \avg{x^{2n}}_{\mathrm{eq}} - x^{2n} \\
    & = - h(x)\,.
\end{align}

Therefore, we obtain
\begin{align}
    I
    & = -\frac{\gamma}{2\lambda_w^\alpha} \int_0^\infty dt\,
    \avg{h \, e^{\mathcal{L}t}\mathcal{L}(x^2)}_{\mathrm{eq}}\\
    & = -\frac{\gamma}{2\lambda_w^\alpha} \int_0^\infty dt\,
    \left\langle
        h \, \frac{d}{dt}\left[e^{\mathcal{L}t}(x^2)\right]
    \right\rangle_{\mathrm{eq}}\\
    & = \frac{\gamma}{2\lambda_w^\alpha} 
    \left[ \avg{h\, x^2}_{\mathrm{eq}} 
    - \lim_{t \to \infty} \avg{h\, e^{\mathcal{L}t}(x^2)}_{\mathrm{eq}}\right]\,.
\end{align}
As $t\rightarrow \infty$, $e^{\mathcal{L}t}(x^2)$ relaxes to the equilibrium average $\avg{x^2}_{\mathrm{eq}}$. Hence, the second term becomes $\avg{h}_{\mathrm{eq}}\avg{x^2}_{\mathrm{eq}} = 0$. Therefore, $I$ can be decomposed into equal-time even moments as
\begin{align}
  I = \frac{\gamma}{2\lambda_w^\alpha}
    \left[
        \avg{x^{2n+2}}_{\mathrm{eq}}
        -
        \avg{x^{2n}}_{\mathrm{eq}}\,
        \avg{x^2}_{\mathrm{eq}}
    \right]\,.
    \label{eq:I-moments}
\end{align}

To evaluate Eq.~(\ref{eq:I-moments}), we use
\begin{equation}
    \avg{x^{2r}}_{\mathrm{eq}}
    =
    \left(
        \frac{2n k_{\mathrm B}T}{\lambda_w^\alpha}
    \right)^{r/n}
    \frac{
        \Gamma\left(\frac{2r+1}{2n}\right)
    }{
        \Gamma\left(\frac{1}{2n}\right)
    }\,,
    \label{eq:x2n-moment}
\end{equation}
from which it follows that
\begin{equation}
  \avg{x^{2n+2}}_{\mathrm{eq}} = 3\frac{k_{\mathrm B}T}{\lambda_w^\alpha}
    \avg{x^2}_{\mathrm{eq}}\,.
    \label{eq:x2nplus2}
\end{equation}
Substituting Eqs.~(\ref{eq:x2n-moment}) and (\ref{eq:x2nplus2}) into Eq.~(\ref{eq:I-moments}) yields
\begin{align}
  I &= \gamma
    \frac{k_{\mathrm B}T}{\lambda_w^{2\alpha}}
    \avg{x^2}_{\mathrm{eq}} \nonumber\\
    &= \gamma \frac{k_{\mathrm B}T}{\lambda_w^{2\alpha}}
    \left(
        \frac{2n k_{\mathrm B}T}{\lambda_w^\alpha}
    \right)^{1/n}
    \frac{
        \Gamma\left(\frac{3}{2n}\right)
    }{
        \Gamma\left(\frac{1}{2n}\right)
    }\,.
    \label{eq:I-final}
\end{align}

For the denominator of Eq.~(\ref{eq:taubar_by_timeintegral}), the equal-time variance $C_{x^{2n}}(0)$ can be written as
\begin{align}
    C_{x^{2n}}(0)
    &=
    \left\langle h^2\right\rangle_{\mathrm{eq}}
    \nonumber\\
    &=
    \left\langle x^{4n}\right\rangle_{\mathrm{eq}}
    -
    \left\langle x^{2n}\right\rangle_{\mathrm{eq}}^2 \nonumber\\
    &= 2n
    \left(
        \frac{k_{\mathrm B}T}{\lambda_w^\alpha}
    \right)^2 .
    \label{eq:C0}
\end{align}

From Eqs.~(\ref{eq:I-final}) and (\ref{eq:C0}), the effective relaxation time can be written as
\begin{align}
    \overline{\tau}_{\mu\nu}
    &= \frac{I}{C_{x^{2n}}(0)} \nonumber\\
    &=
    \frac{f_n}{2n} \gamma\, (k_{\mathrm B}T)^{-1+\frac{1}{n}}
    \lambda_w^{-\frac{\alpha}{n}},
\end{align}
where
\begin{equation}
    f_n \equiv 
    (2n)^{\frac{1}{n}}\,
    \frac{
        \Gamma\left(\frac{3}{2n}\right)
    }{
        \Gamma\left(\frac{1}{2n}\right)
    }\,.
    \label{eq:fn-final}
\end{equation}

\section{Expressions for $C_{\mu\nu}^{(i)}$ in Eq.~(\ref{eq:correxp_udb})}\label{app:udb}

The equilibrium two-time correlation functions of the generalized forces given by Eqs.~(\ref{eq:dxwdxw_udb})--(\ref{eq:dxudxu_udb}) can be expressed as
\begin{align}
  \avg{\Delta X_\mu(t)\, \Delta X_\nu(t')}_{\rm eq}
  = \sigma_{\mu\nu}(t)\, \sum_{i=1}^{3} C_{\mu\nu}^{(i)}\, e^{-|t-t'|/\tau_{\mu\nu}^{(i)}}\,,
\end{align}
where $\tau_{\mu\nu}^{(i)}$ ($i=1$--$3$) are the time constants defined in Eqs.~(\ref{eq:tau1_udb})--(\ref{eq:tau3_udb}). Below, we list the explicit expressions for the coefficients $C_{\mu\nu}^{(i)}$ for each $(\mu,\,\nu)$ component.

\begin{itemize}
 \item[(i)] $(w,w)$ component:
\begin{align}
  C_{ww}^{(1)} =& \frac{\Gamma_+^2}{(\Gamma_+ - \Gamma_-)^2}\,,\\
  C_{ww}^{(2)} =& \frac{-2\Gamma_+\Gamma_-}{(\Gamma_+ - \Gamma_-)^2}\,,\\
  C_{ww}^{(3)} =& \frac{\Gamma_-^2}{(\Gamma_+ - \Gamma_-)^2}\,.
\end{align}
\item[(ii)] $(w,u)$ component:
\begin{align}
  C_{wu}^{(1)} =& \frac{1}{(\Gamma_+ - \Gamma_-)^2} \left(\Gamma_+^2 + \frac{\lambda_w}{M}\right)\,,\\
  C_{wu}^{(2)} =& \frac{-2}{(\Gamma_+ - \Gamma_-)^2} \left(\Gamma_+\Gamma_- + \frac{\lambda_w}{M}\right)\,,\\
  C_{wu}^{(3)} =& \frac{1}{(\Gamma_+ - \Gamma_-)^2} \left(\Gamma_-^2 + \frac{\lambda_w}{M}\right)\,.
\end{align}
\item[(iii)] $(u,w)$ component:
\begin{align}
  C_{uw}^{(1)} =& \frac{\Gamma_+^2}{(\Gamma_+ - \Gamma_-)^2} \left(1 + \frac{M}{\lambda_w} \Gamma_-^2\right)\,,\\
  C_{uw}^{(2)} =& \frac{-2 \Gamma_+\Gamma_-}{(\Gamma_+ - \Gamma_-)^2} \left(1 + \frac{M}{\lambda_w} \Gamma_+\Gamma_-\right)\,,\\
  C_{uw}^{(3)} =& \frac{\Gamma_-^2}{(\Gamma_+ - \Gamma_-)^2} \left(1 + \frac{M}{\lambda_w} \Gamma_+^2\right)\,.
\end{align}
\item[(iv)] $(u,u)$ component:
\begin{align}
  C_{uu}^{(1)} =& \frac{1}{2(\Gamma_+ - \Gamma_-)^2} \left(\Gamma_+^2 + \Gamma_-^2 + \frac{M}{\lambda_w} \Gamma_+^2\Gamma_-^2 + \frac{\lambda_w}{M} \right)\,,\\
  C_{uu}^{(2)} =& \frac{-1}{(\Gamma_+ - \Gamma_-)^2} \left(2 \Gamma_+\Gamma_- + \frac{M}{\lambda_w} \Gamma_+^2\Gamma_-^2 + \frac{\lambda_w}{M} \right)\,,\\
  C_{uu}^{(3)} =& \frac{1}{2(\Gamma_+ - \Gamma_-)^2} \left(\Gamma_+^2 + \Gamma_-^2 + \frac{M}{\lambda_w} \Gamma_+^2\Gamma_-^2 + \frac{\lambda_w}{M} \right)\,.
\end{align}
\end{itemize}

\section{Equations of motion for the first and the second moments for an underdamped Brownian particle in a harmonic potential}\label{app:eoms}

In this appendix, we summarize the equations of motion for the first and second moments of the position $x$ and momentum $p$ used in the numerical analysis of Sec.~\ref{sec:brownian_carnot}. These equations follow from the Fokker-Planck equation for an underdamped Brownian particle in a 1D harmonic potential $V_{\lambda_w}$ given by Eq.~(\ref{eq:vho}):
\begin{equation}
  \frac{\partial}{\partial t} \rho + \frac{p}{M} \frac{\partial}{\partial x} \rho - \lambda_w(t) x \frac{\partial}{\partial p} \rho - \frac{\gamma}{M} \frac{\partial}{\partial p}(p\rho) - \frac{\gamma}{\beta(t)} \frac{\partial^2}{\partial p^2} \rho = 0\,,\label{eq:fp_udb_app}
\end{equation}
where $\lambda_w(t)$ and $\beta(t) \equiv 1/k_{\mathrm{B}}T(t)$ serve as time-dependent control parameters. [Equation~(\ref{eq:fp_udb_app}) is identical to Eq.~(\ref{eq:fp_udb}) and is reproduced here for the readers' convenience.]

For a harmonic potential, the phase-space distribution function $\rho(x,p,t)$ remains Gaussian at all times provided the initial distribution is Gaussian. Consequently, the dynamics is fully characterized by the first and second moments of $x$ and $p$. Using Eq.~(\ref{eq:fp_udb_app}) and integrating by parts, one readily obtains equations of motion for these moments. The first moments $\avg{x}$ and $\avg{p}$ obey a closed set of linear differential equations:
\begin{align}
  \frac{d}{dt}\avg{x} &= \frac{1}{M}\avg{p}\,,\\
  \frac{d}{dt}\avg{p} &= -\lambda_w(t)\,\avg{x} - \frac{\gamma}{M}\avg{p}\,.
\end{align}
The second moments $\avg{x^2}$, $\avg{xp}$, and $\avg{p^2}$ obey another closed set of linear differential equations:
\begin{align}
  \frac{d}{dt}\avg{x^2} &= \frac{2}{M} \avg{xp}\,,\\
  \frac{d}{dt}\avg{xp} &= \frac{1}{M} \avg{p^2} - \lambda_w(t)\, \avg{x^2} - \frac{\gamma}{M} \avg{xp}\,,\\
  \frac{d}{dt}\avg{p^2} &= -2\lambda_w(t) \avg{xp} - 2\frac{\gamma}{M} \avg{p^2} + 2\gamma k_{\mathrm{B}}T(t)\,.
\end{align}
These equations are numerically integrated for the protocols $\lambda_w(t)$ and $T(t)$ specified in Sec.~\ref{sec:brownian_carnot}. Typically, after one or two cycles, the phase-space distribution relaxes to a periodic state satisfying $\rho(x,p,t) = \rho(x,p,t+\tau)$. Thermodynamic quantities of the cycle are evaluated after the system has reached this periodic state. For the present harmonic system, this moment-based approach is equivalent to solving the full Fokker-Planck equation, while being numerically more efficient and transparent.

The average values of $W$, $U$, and $A$, as well as the power and efficiency, are evaluated by integrating the corresponding time-dependent moments over one cycle. In particular, $\avg{W}$ and $\avg{U}$ are given by
\begin{align}
  \avg{W} &= -\int_0^\tau dt\, \left\langle \frac{\partial H_{\lambda_w}}{\partial \lambda_w} \right\rangle \dot{\lambda}_w = -\frac{1}{2} \int_0^\tau dt\, \avg{x^2}\, \dot{\lambda}_w\,,\\
  \avg{U} &= -\int_0^\tau dt\, S \dot{T}\nonumber\\
  &= -k_{\mathrm{B}} \int_0^\tau dt\, \left[ \ln{2\pi} + 1 + \frac{1}{2} \ln{\left(\avg{x^2}\avg{p^2} - \avg{xp}^2 \right)} \right] \dot{T}\,.
\end{align}
Here, to express $\avg{U}$ in terms of the second moments, we have used the fact that a Gaussian phase-space distribution can be written as
\begin{align}
  &\rho(x, p) \nonumber\\
  &= \frac{1}{2\pi\sqrt{\avg{x^2}\avg{p^2} - \avg{xp}^2}}\, \exp{\left[ -\frac{\avg{p^2} x^2 + \avg{x^2} p^2 - 2 \avg{xp} xp}{2 \left(\avg{x^2}\avg{p^2} - \avg{xp}^2 \right)} \right]}\,.
\end{align}
While the mean values of thermodynamic quantities, which are determined by one-time moments, are evaluated by solving the full Fokker-Planck equation as explained above, the fluctuation of the dissipation $\avg{\Delta A^2}$ and the fluctuation of the stochastic efficiency $\avg{\Delta\mathcal{E}^2}$ are evaluated within the linear-response approximation from Eqs.~(\ref{eq:vara}) and (\ref{eq:var_efficiency}), respectively.


\begin{thebibliography}{99}

\bibitem{Bustamante05} C. Bustamante, J. Liphardt, and F. Ritort, The Nonequilibrium Thermodynamics of Small Systems, \href{https://doi.org/10.1063/1.2012462}{Phys.\ Today {\bf 58}, 43 (2005)}.

\bibitem{Ciliberto17} S. Ciliberto, Experiments in Stochastic Thermodynamics: Short History and Perspectives, \href{https://doi.org/10.1103/PhysRevX.7.021051}{Phys.\ Rev.\ X {\bf 7}, 021051 (2017)}.


\bibitem{Blickle12} V. Blickle and C. Bechinger, Realization of a micrometre-sized stochastic heat engine, \href{https://doi.org/10.1038/nphys2163}{Nat.\ Phys. {\bf 8}, 143 (2012)}.

\bibitem{Quinto-Su14} P.~A. Quinto-Su, A microscopic steam engine implemented in an optical tweezer, \href{https://doi.org/10.1038/ncomms6889}{Nat.\ Commun. {\bf 5}, 5889 (2014)}.

\bibitem{Martinez16} I.~A. Mart\'inez, \'E. Rold\'an, L. Dinis, D. Petrov, J.~M.~R. Parrondo, and R.~A. Rica, Brownian Carnot engine, \href{https://doi.org/10.1038/nphys3518}{Nat.\ Phys. {\bf 12}, 67 (2016)}.

\bibitem{Rossnagel16} J. Ro{\ss}nagel, S.~T. Dawkins, K.~N. Tolazzi, O. Abah, E. Lutz, F. Schmidt-Kaler, and K. Singer, A single-atom heat engine, \href{https://doi.org/10.1126/science.aad6320}{Science {\bf 352}, 325 (2016)}.

\bibitem{Martinez17} I.~A. Mart\'inez, \'E. Rold\'an, L. Dinis, and R.~A. Rica, Colloidal heat engines: a review, \href{https://doi.org/10.1039/C6SM00923A}{Soft Matter, {\bf 13}, 22 (2017)}.

\bibitem{Argun17} A. Argun, J. Soni, L. Dabelow, S. Bo, G. Pesce, R. Eichhorn, and G. Volpe, Experimental realization of a minimal microscopic heat engine, \href{https://doi.org/10.1103/PhysRevE.96.052106}{Phys.\ Rev.\ E {\bf 96}, 052106 (2017)}.

\bibitem{Maslennikov19} G. Maslennikov, S. Ding, R. Habl\"utzel, J. Gan, A. Roulet, S. Nimmrichter, J. Dai, V. Scarani, and D. Matsukevich, Quantum absorption refrigerator with trapped ions, \href{https://doi.org/10.1038/s41467-018-08090-0}{Nat.\ Commun. {\bf 10}, 202 (2019)}.

\bibitem{vonLindenfels19} D. von Lindenfels, O. Gr\"ab, C.~T. Schmiegelow, V. Kaushal, J. Schulz, M.~T. Mitchison, J. Goold, F. Schmidt-Kaler, and U.~G. Poschinger, Spin Heat Engine Coupled to a Harmonic-Oscillator Flywheel, \href{https://doi.org/10.1103/PhysRevLett.123.080602}{Phys.\ Rev.\ Lett. {\bf 123}, 080602 (2019)}.

\bibitem{Hou25} W. Hou, W. Yao, X. Zhao, K. Rehan, Y. Li, Y. Li, E. Lutz, Y. Lin, J. Du, Combining energy efficiency and quantum advantage in cyclic machines, \href{https://doi.org/10.1038/s41467-025-60179-5}{Nat.\ Commun. {\bf 16}, 5127 (2025)}.

\bibitem{Steeneken11} P.~G. Steeneken, K. Le Phan, M.~J. Goossens, G.~E.~J. Koops, G.~J.~A.~M. Brom, C. van der Avoort, and J.~T.~M. van Beek, Piezoresistive heat engine and refrigerator, \href{https://doi.org/10.1038/nphys1871}{Nat.\ Phys. {\bf 7}, 354 (2011)}.

\bibitem{Pekola15} J.~P. Pekola, Towards quantum thermodynamics in electronic circuits, \href{https://doi.org/10.1038/nphys3169}{Nat.\ Phys. {\bf 11}, 118 (2015)}.

\bibitem{Hugel02} T. Hugel, N.~B. Holland, A. Cattani, L. Moroder, M. Seitz, H.~E. Gaub, Single-Molecule Optomechanical Cycle, \href{https://doi.org/10.1126/science.1069856}{Science {\bf 296}, 1103 (2002)}.

\bibitem{Klaers17} J. Klaers, S. Faelt, A. Imamoglu, and E. Togan, Squeezed Thermal Reservoirs as a Resource for a Nanomechanical Engine beyond the Carnot Limit, \href{https://doi.org/10.1103/PhysRevX.7.031044}{Phys.\ Rev.\ X {\bf 7}, 031044 (2017)}.


\bibitem{Seifert12} U. Seifert, Stochastic thermodynamics, fluctuation theorems and molecular machines, \href{https://doi.org/10.1088/0034-4885/75/12/126001}{Rep.\ Prog.\ Phys. {\bf 75}, 126001 (2012)}.


\bibitem{Sekimoto98} K. Sekimoto, Langevin Equation and Thermodynamics, \href{https://doi.org/10.1143/PTPS.130.17}{Prog.\ Theor.\ Phys.\ Suppl. {\bf 130}, 17 (1998)}.

\bibitem{Sekimotobook10} K. Sekimoto, \href{https://doi.org/10.1007/978-3-642-05411-2}{{\it Stochastic Energetics}, Lecture Notes in Physics (Springer, Berlin, 2010), Vol. 799}.

\bibitem{Saitobook22} K. Saito, \href{https://www.saiensu.co.jp/search/?isbn=978-4-7819-1563-0&y=2022}{{\it Thermodynamics of Fluctuating Systems---Development of Non-Equilibrium Statistical Mechanics} (in Japanese), SGC Library (Saiensu-Sha, Tokyo, 2022), Vol. 182}.

\bibitem{Shiraishibook23} N. Shiraishi, \href{https://doi.org/10.1007/978-981-19-8186-9}{{\it An Introduction to Stochastic Thermodynamics---From Basic to Advanced}, (Springer, Singapore, 2023)}.

\bibitem{Pelitibook21} L. Peliti and S. Pigolotti, \href{https://press.princeton.edu/books/hardcover/9780691201771/stochastic-thermodynamics?srsltid=AfmBOoq3QjxtlxpZr-DYxq1i--fNjfi8k7GzkhT3A9n9rasOwu54G9G2}{{\it Stochastic Thermodynamics: An Introduction}, (Princeton University Press, Princeton, 2021)}.

\bibitem{Seifertbook25} U. Seifert, \href{https://doi.org/10.1017/9781009024358}{{\it Stochastic Thermodynamics}, (Cambridge University Press, Cambridge, 2025)}.

\bibitem{Jarzynski11} C. Jarzynski, Equalities and Inequalities: Irreversibility and the Second Law of Thermodynamics at the Nanoscale, \href{https://doi.org/10.1146/annurev-conmatphys-062910-140506}{Ann.\ Rev.\ Cond.\ Mat.\ Phys. {\bf 2}, 329 (2011)}.

\bibitem{VandenBroeck14} C. Van den Broeck and M. Esposito, Ensemble and trajectory thermodynamics: A brief introduction, \href{https://doi.org/10.1016/j.physa.2014.04.035}{Physica A {\bf 418}, 6 (2015)}.

\bibitem{Nicolis17} G. Nicolis and Y. De Decker, Stochastic Thermodynamics of Brownian Motion, \href{https://doi.org/10.3390/e19090434}{Entropy {\bf 19}, 434 (2017)}.

\bibitem{Guery-Odelin23} D. Gu\'ery-Odelin, C. Jarzynski, C.~A. Plata, A. Prados, and E. Trizac, Driving rapidly while remaining in control: classical shortcuts from Hamiltonian to stochastic dynamics, \href{https://doi.org/10.1088/1361-6633/acacad}{Rep.\ Prog.\ Phys. {\bf 86}, 035902 (2023)}.







\bibitem{Krishnamurthy16} S. Krishnamurthy, S. Ghosh, D. Chatterji, R. Ganapathy, and A.~K. Sood, A micrometre-sized heat engine operating between bacterial reservoirs, \href{https://doi.org/10.1038/nphys3870}{Nat.\ Phys. {\bf 12}, 1134 (2016)}.

\bibitem{Pietzonka19} P. Pietzonka, \'E. Fodor, C. Lohrmann, M.~E. Cates, and U. Seifert, Autonomous Engines Driven by Active Matter: Energetics and Design Principles, \href{https://doi.org/10.1103/PhysRevX.9.041032}{Phys.\ Rev.\ X {\bf 9}, 041032 (2019)}.

\bibitem{Fodor21} \'E. Fodor and M.~E. Cates, Active engines: Thermodynamics moves forward, \href{https://doi.org/10.1209/0295-5075/134/10003}{Europhys.\ Lett. {\bf 134}, 10003 (2021)}.

\bibitem{Toyabe15} S. Toyabe and M. Sato, Nonequilibrium Fluctuations in Biological Strands, Machines, and Cells, \href{https://doi.org/10.7566/JPSJ.84.102001}{J.\ Phys.\ Soc.\ Jpn. {\bf 84}, 102001 (2015)}.


\bibitem{Jarzynski97} C. Jarzynski, Nonequilibrium Equality for Free Energy Differences, \href{https://doi.org/10.1103/PhysRevLett.78.2690}{Phys.\ Rev.\ Lett. {\bf 78}, 2690 (1997)}.

\bibitem{Crooks99} G.~E. Crooks, Entropy production fluctuation theorem and the nonequilibrium work relation for free energy differences, \href{https://doi.org/10.1103/PhysRevE.60.2721}{Phys.\ Rev.\ E {\bf 60}, 2721 (1999)}.


\bibitem{Barato15} A.~C. Barato and U. Seifert, Thermodynamic Uncertainty Relation for Biomolecular Processes, \href{https://doi.org/10.1103/PhysRevLett.114.158101}{Phys.\ Rev.\ Lett. {\bf 114}, 158101 (2015)}.

\bibitem{Gingrich16} T.~R. Gingrich, J.~M. Horowitz, N. Perunov, and J.~L. England, Dissipation Bounds All Steady-State Current Fluctuations, \href{https://doi.org/10.1103/PhysRevLett.116.120601}{Phys.\ Rev.\ Lett. {\bf 116}, 120601 (2016)}.

\bibitem{Seifert19} U. Seifert, From Stochastic Thermodynamics to Thermodynamic Inference, \href{https://doi.org/10.1146/annurev-conmatphys-031218-013554}{Annu.\ Rev.\ Condens.\ Matter Phys. {\bf 10}, 171 (2019)}.

\bibitem{Hasegawa19} Y. Hasegawa and T. Van Vu, Fluctuation Theorem Uncertainty Relation, \href{https://doi.org/10.1103/PhysRevLett.123.110602}{Phys.\ Rev.\ Lett. {\bf 123}, 110602 (2019)}.

\bibitem{Horowitz20} J. M. Horowitz and T. R. Gingrich, Thermodynamic uncertainty relations constrain non-equilibrium fluctuations, \href{https://www.nature.com/articles/s41567-019-0702-6}{Nat.\ Phys. {\bf 16}, 15 (2020)}.

\bibitem{Liu20} K. Liu, Z. Gong, and M. Ueda, Thermodynamic Uncertainty Relation for Arbitrary Initial States, \href{https://doi.org/10.1103/PhysRevLett.125.140602}{Phys.\ Rev.\ Lett. {\bf 125}, 140602 (2020)}.

\bibitem{Koyuk20} T. Koyuk and U. Seifert, Thermodynamic Uncertainty Relation for Time-Dependent Driving, \href{https://doi.org/10.1103/PhysRevLett.125.260604}{Phys.\ Rev.\ Lett. {\bf 125}, 260604 (2020)}.

\bibitem{Tan20} T. Van Vu and Y. Hasegawa, Thermodynamic uncertainty relations under arbitrary control protocols, \href{https://doi.org/10.1103/PhysRevResearch.2.013060}{Phys.\ Rev.\ Research {\bf 2}, 013060 (2020)}.

\bibitem{Barato18} A.~C. Barato, R. Chetrite, A. Faggionato, and D. Gabrielli, Bounds on current fluctuations in periodically driven systems, \href{https://doi.org/10.1088/1367-2630/aae512}{New J.\ Phys. {\bf 20}, 103023 (2018)}. 

\bibitem{Koyuk19a} T. Koyuk, U. Seifert, and P. Pietzonka, A generalization of the thermodynamic uncertainty relation to periodically driven systems, \href{https://doi.org/10.1088/1751-8121/aaeec4}{J.\ Phys.\ A: Math.\ Theor. {\bf 52}, 02LT02 (2019)}. 

\bibitem{Koyuk19b} T. Koyuk and U. Seifert, Operationally Accessible Bounds on Fluctuations and Entropy Production in Periodically Driven Systems, \href{https://doi.org/10.1103/PhysRevLett.122.230601}{Phys.\ Rev.\ Lett. {\bf 122}, 230601 (2019)}. 

\bibitem{Miller21} H.~J.~D. Miller, M. H. Mohammady, M. Perarnau-Llobet, and G. Guarnieri, Thermodynamic Uncertainty Relation in Slowly Driven Quantum Heat Engines, \href{https://doi.org/10.1103/PhysRevLett.126.210603}{Phys.\ Rev.\ Lett. {\bf 126}, 210603 (2021)}.

\bibitem{Wang24} Z. Wang and J. Ren, Thermodynamic Geometry of Nonequilibrium Fluctuations in Cyclically Driven Transport, \href{https://doi.org/10.1103/PhysRevLett.132.207101}{Phys.\ Rev.\ Lett. {\bf 132}, 207101 (2024)}.



\bibitem{Sekimoto00} K. Sekimoto, F. Takagi, and T. Hondou, Carnot's cycle for small systems: Irreversibility and cost of operations, \href{https://doi.org/10.1103/PhysRevE.62.7759}{Phys.\ Rev.\ E {\bf 62}, 7759 (2000)}.

\bibitem{VandenBroeck05} C. Van den Broeck, Thermodynamic Efficiency at Maximum Power, \href{https://doi.org/10.1103/PhysRevLett.95.190602}{Phys.\ Rev.\ Lett {\bf 95}, 190602 (2005)}.

\bibitem{Schmiedl08} T. Schmiedl and U. Seifert, Efficiency at maximum power: An analytically solvable model for stochastic heat engines, \href{https://doi.org/10.1209/0295-5075/81/20003}{Europhys. Lett. {\bf 81}, 200003 (2008)}.

\bibitem{Esposito10} M. Esposito, R. Kawai, K. Lindenberg, and C. Van den Broeck, Efficiency at Maximum Power of Low-Dissipation Carnot Engines, \href{https://doi.org/10.1103/PhysRevLett.105.150603}{Phys.\ Rev.\ Lett. {\bf 105}, 150603 (2010)}.

\bibitem{Dechant15} A. Dechant, N. Kiesel, and E. Lutz, All-Optical Nanomechanical Heat Engine, \href{https://doi.org/10.1103/PhysRevLett.114.183602}{Phys. Rev. Lett. {\bf 114}, 183602 (2015)}.

\bibitem{Brandner15} K. Brandner, K. Saito, and U. Seifert, Thermodynamics of Micro- and Nano-Systems Driven by Periodic Temperature Variations, \href{https://doi.org/10.1103/PhysRevX.5.031019}{Phys.\ Rev.\ X {\bf 5}, 031019 (2015)}.

\bibitem{Dechant17} A. Dechant, N. Kiesel and E. Lutz, Underdamped stochastic heat engine at maximum efficiency, \href{https://doi.org/10.1209/0295-5075/119/50003}{Europhys.\ Lett. {\bf 119}, 50003 (2017)}.



\bibitem{Strasberg21} P. Strasberg, C.~W. W\"achtler, and G. Schaller, Autonomous Implementation of Thermodynamic Cycles at the Nanoscale, \href{https://doi.org/10.1103/PhysRevLett.126.180605}{Phys.\ Rev.\ Lett. {\bf 126}, 180605 (2021)}.




\bibitem{Hoppenau13} J. Hoppenau, M. Niemann, and A. Engel, Carnot process with a single particle, \href{https://doi.org/10.1103/PhysRevE.87.062127}{Phys.\ Rev.\ E {\bf 87}, 062127 (2013)}.

\bibitem{Holubec14} V. Holubec, An exactly solvable model of a stochastic heat engine: optimization of power, power fluctuations and efficiency, \href{https://doi.org/10.1088/1742-5468/2014/05/P05022}{J.\ Stat.\ Mech. {\bf 2014}, P05022 (2014)}.

\bibitem{Rana14} S. Rana, P.~S. Pal, A. Saha, and A.~M. Jayannavar, Single-particle stochastic heat engine, \href{https://doi.org/10.1103/PhysRevE.90.042146}{Phys.\ Rev.\ E {\bf 90}, 042146 (2014)}.

\bibitem{Cerino15} L. Cerino, A. Puglisi, and A. Vulpiani, Kinetic model for the finite-time thermodynamics of small heat engines, \href{https://doi.org/10.1103/PhysRevE.91.032128}{Phys.\ Rev.\ E {\bf 91}, 032128 (2015)}.

\bibitem{Holubec17} V. Holubec and A. Ryabov, Work and power fluctuations in a critical heat engine, \href{https://doi.org/10.1103/PhysRevE.96.030102}{Phys.\ Rev.\ E {\bf 96}, 030102(R) (2017)}.

\bibitem{Holubec18} V. Holubec and A. Ryabov, Cycling Tames Power Fluctuations near Optimum Efficiency, \href{https://doi.org/10.1103/PhysRevLett.121.120601}{Phys.\ Rev.\ Lett. {\bf 121}, 120601 (2018)}.

\bibitem{Dechant19} A. Dechant, Multidimensional thermodynamic uncertainty relations, \href{https://doi.org/10.1088/1751-8121/aaf3ff}{J.\ Phys.\ A: Math.\ Theor. {\bf 52}, 035001 (2019)}.

\bibitem{Ito25} K. Ito, G.-H. Xu, C. Jiang, \'E. Rold\'an, R.~A. Rica-Alarc\'on, I.~A. Mart\'inez, and G. Watanabe, Universal relations and bounds for fluctuations in quasistatic small heat engines, \href{https://doi.org/10.1038/s42005-025-01961-1}{Commun.\ Phys. {\bf 8}, 60 (2025)}.

\bibitem{Saryal21} S. Saryal, M. Gerry, I. Khait, D. Segal, and B.~K. Agarwalla, Universal Bounds on Fluctuations in Continuous Thermal Machines, \href{https://doi.org/10.1103/PhysRevLett.127.190603}{Phys.\ Rev.\ Lett. {\bf 127}, 190603 (2021)}.

\bibitem{Mohanta21} S. Mohanta, S. Saryal, and B.~K. Agarwalla, Universal bounds on cooling power and cooling efficiency for autonomous absorption refrigerators, \href{https://arxiv.org/abs/2106.12809}{arXiv:2106.12809 [cond-mat.stat-mech] (2021)}.

\bibitem{Holubec21} V. Holubec and A. Ryabov, Fluctuations in heat engines, \href{https://doi.org/10.1088/1751-8121/ac3aac}{J.\ Phys.\ A: Math.\ Theor. {\bf 55}, 013001 (2021)}.



\bibitem{Verley14a} G. Verley, M. Esposito, T. Willaert, and C. Van den Broeck, The unlikely Carnot efficiency, \href{https://doi.org/10.1038/ncomms5721}{Nat.\ Commun. {\bf 5}, 4721 (2014)}.

\bibitem{Verley14b} G. Verley, T. Willaert, C. Van den Broeck, and M. Esposito, Universal theory of efficiency fluctuations, \href{https://doi.org/10.1103/PhysRevE.90.052145}{Phys.\ Rev.\ E {\bf 90}, 052145 (2014)}.

\bibitem{Proesmans15a} K. Proesmans and C. Van den Broeck, Stochastic efficiency: five case studies, \href{https://doi.org/10.1088/1367-2630/17/6/065004}{New.\ J.\ Phys. {\bf 17}, 065004 (2015)}.

\bibitem{Proesmans15b} K. Proesmans, C. Driesen, B. Cleuren, and C. Van den Broeck, Efficiency of single-particle engines, \href{https://doi.org/10.1103/PhysRevE.92.032105}{Phys.\ Rev.\ E {\bf 92}, 032105 (2015)}.

\bibitem{Park16} J.-M. Park, H.-M. Chun, and J.~D. Noh, Efficiency at maximum power and efficiency fluctuations in a linear Brownian heat-engine model, \href{https://doi.org/10.1103/PhysRevE.94.012127}{Phys.\ Rev.\ E {\bf 94}, 012127 (2016)}.

\bibitem{Saha18} A. Saha, R. Marathe, P.~S. Pal, and A.~M. Jayannavar, Stochastic heat engine powered by active dissipation, \href{https://doi.org/10.1088/1742-5468/aae84a}{J.\ Stat.\ Mech. {\bf 2018}, 113203 (2018)}.

\bibitem{Manikandan19} S.~K. Manikandan, L. Dabelow, R. Eichhorn, and S. Krishnamurthy, Efficiency Fluctuations in Microscopic Machines, \href{https://doi.org/10.1103/PhysRevLett.122.140601}{Phys.\ Rev.\ Lett. {\bf 122}, 140601 (2019)}.

\bibitem{Vroylandt20} H. Vroylandt, M. Esposito, and G. Verley, Efficiency Fluctuations of Stochastic Machines Undergoing a Phase Transition, \href{https://doi.org/10.1103/PhysRevLett.124.250603}{Phys.\ Rev.\ Lett. {\bf 124}, 250603 (2020)}.



\bibitem{Sinitsyn11} N.~A. Sinitsyn, Fluctuation relation for heat engines, \href{https://doi.org/10.1088/1751-8113/44/40/405001}{J. Phys.\ A {\bf 44}, 405001 (2011)}.

\bibitem{Pal17} P.~S. Pal, S. Lahiri, and A.~M. Jayannavar, Transient exchange fluctuation theorems for heat using a Hamiltonian framework: Classical and quantum regimes, \href{https://doi.org/10.1103/PhysRevE.95.042124}{Phys.\ Rev.\ E {\bf 95}, 042124 (2017)}.



\bibitem{Proesmans17} K. Proesmans and C. Van den Broeck, Discrete-time thermodynamic uncertainty relation, \href{https://doi.org/10.1209/0295-5075/119/20001}{Europhys.\ Lett. {\bf 119}, 20001 (2017)}. 

\bibitem{Pietzonka18} P. Pietzonka and U. Seifert, Universal Trade-Off between Power, Efficiency, and Constancy in Steady-State Heat Engines, \href{https://doi.org/10.1103/PhysRevLett.120.190602}{Phys.\ Rev.\ Lett. {\bf 120}, 190602 (2018)}. 

\bibitem{Timpanaro19} A.~M. Timpanaro, G. Guarnieri, J. Goold, and G.~T. Landi, Thermodynamic Uncertainty Relations from Exchange Fluctuation Theorems, \href{https://doi.org/10.1103/PhysRevLett.123.090604}{Phys.\ Rev.\ Lett. {\bf 123}, 090604 (2019)}. 

\bibitem{Kamijima21} T. Kamijima, S. Otsubo, Y. Ashida, and T. Sagawa, Higher-order efficiency bound and its application to nonlinear nanothermoelectrics, \href{https://doi.org/10.1103/PhysRevE.104.044115}{Phys.\ Rev.\ E {\bf 104}, 044115 (2021)}.

\bibitem{Xu21} G.-H. Xu and G. Watanabe, Correlation-enhanced stability of microscopic cyclic heat engines, \href{https://doi.org/10.1103/PhysRevResearch.4.L032017}{Phys.\ Rev.\ Research {\bf 4}, L032017 (2022)}.

\bibitem{Xu22} G.-H. Xu, C. Jiang, Y. Minami, and G. Watanabe, Relation between fluctuations and efficiency at maximum power for small heat engines, \href{https://doi.org/10.1103/PhysRevResearch.4.043139}{Phys.\ Rev.\ Research {\bf 4}, 043139 (2022)}.


\bibitem{Plata19} C.~A. Plata, D. Gu\'ery-Odelin, E. Trizac, and A. Prados, Optimal work in a harmonic trap with bounded stiffness, \href{https://doi.org/10.1103/PhysRevE.99.012140}{Phys.\ Rev.\ E {\bf 99}, 012140 (2019)}.

\bibitem{Plata20a} C.~A. Plata, D. Gu\'ery-Odelin, E. Trizac, and A. Prados, Building an irreversible Carnot-like heat engine with an overdamped harmonic oscillator, \href{https://doi.org/10.1088/1742-5468/abb0e1}{J.\ Stat.\ Mech. {\bf 2020}, 093207 (2020)}. 

\bibitem{Plata20b} C.~A. Plata, D. Gu\'ery-Odelin, E. Trizac, and A. Prados, Finite-time adiabatic processes: Derivation and speed limit, \href{https://doi.org/10.1103/PhysRevE.101.032129}{Phys.\ Rev.\ E {\bf 101}, 032129 (2020)}. 




\bibitem{Jop08} P. Jop, A. Petrosyan, and S. Ciliberto, Work and dissipation fluctuations near the stochastic resonance of a colloidal particle, \href{https://doi.org/10.1209/0295-5075/81/50005}{Europhys.\ Lett. {\bf 81}, 50005 (2008)}.

\bibitem{Martinez15} I.~A. Mart\'inez, \'E. Rold\'an, L. Dinis, D. Petrov, and R.~A. Rica, Adiabatic Processes Realized with a Trapped Brownian Particle, \href{https://doi.org/10.1103/PhysRevLett.114.120601}{Phys.\ Rev.\ Lett. {\bf 114}, 120601 (2015)}.




\bibitem{Weinhold75} F. Weinhold, Metric geometry of equilibrium thermodynamics, \href{https://doi.org/10.1063/1.431689}{J.\ Chem.\ Phys. {\bf 63}, 2479 (1975)}. 

\bibitem{Weinhold76} F. Weinhold, Thermodynamics and geometry, \href{https://doi.org/10.1063/1.3023366}{Phys.\ Today {\bf 29}, 23 (1976)}.

\bibitem{Ruppeiner79} G. Ruppeiner, Thermodynamics: A Riemannian geometric model, \href{https://doi.org/10.1103/PhysRevA.20.1608}{Phys.\ Rev.\ A {\bf 20}, 1608 (1979)}. 

\bibitem{Ruppeiner95} G. Ruppeiner, Riemannian geometry in thermodynamic fluctuation theory, \href{https://doi.org/10.1103/RevModPhys.67.605}{Rev.\ Mod.\ Phys. {\bf 67}, 605 (1995)}. 

\bibitem{Andresen11} B. Andresen, Current trends in finite-time thermodynamics,
\href{https://doi.org/10.1002/anie.201001411}{Angew.\ Chem.\ Int.\ Ed. {\bf 50}, 2690 (2011)}.




\bibitem{Salamon83} P. Salamon and R.~S. Berry, Thermodynamic Length and Dissipated Availability, \href{https://doi.org/10.1103/PhysRevLett.51.1127}{Phys.\ Rev.\ Lett. {\bf 51}, 1127 (1983)}.

\bibitem{Andresen84} B. Andresen, P. Salamon, and R.~S. Berry, Thermodynamics in finite time, \href{https://doi.org/10.1063/1.2916405}{Phys.\ Today {\bf 37}, No.~9, 62 (1984)}. 

\bibitem{Nulton85} J. Nulton, P. Salamon, B. Andresen, and Q. Anmin, Quasistatic processes as step equilibrations, \href{https://doi.org/10.1063/1.449774}{J.\ Chem.\ Phys. {\bf 83}, 334 (1985)}. 


\bibitem{Salamon84} P. Salamon, J. Nulton, and E. Ihrig, On the relation between entropy and energy versions of thermodynamic length, \href{https://doi.org/10.1063/1.446467}{J.\ Chem.\ Phys. {\bf 80}, 436 (1984)}. 

\bibitem{Gilmore84} R. Gilmore, Length and curvature in the geometry of thermodynamics, \href{https://doi.org/10.1103/PhysRevA.30.1994}{Phys.\ Rev.\ A {\bf 30}, 1994 (1984)}. 

\bibitem{Schlogl85} F. Schl\"ogl, Thermodynamic metric and stochastic measures, \href{https://doi.org/10.1007/BF01328857}{Z.\ Phys.\ B {\bf 59}, 449 (1985)}. 

\bibitem{Brody94} D. Brody and N. Rivier, Geometrical aspects of statistical mechanics, \href{https://doi.org/10.1103/PhysRevE.51.1006}{Phys.\ Rev.\ E {\bf 51}, 1006 (1995)}. 



\bibitem{Crooks07} G.~E. Crooks, Measuring Thermodynamic Length, \href{https://doi.org/10.1103/PhysRevLett.99.100602}{Phys.\ Rev.\ Lett. {\bf 99}, 100602 (2007)}. 

\bibitem{Zulkowski12} P.~R. Zulkowski, D.~A. Sivak, G.~E. Crooks, and M.~R. DeWeese, Geometry of thermodynamic control, \href{https://doi.org/10.1103/PhysRevE.86.041148}{Phys.\ Rev.\ E {\bf 86}, 041148 (2012)}. 

\bibitem{Vu21} T. Van Vu and Y. Hasegawa, Geometrical Bounds of the Irreversibility in Markovian Systems, \href{https://doi.org/10.1103/PhysRevLett.126.010601}{Phys.\ Rev.\ Lett. {\bf 126}, 010601 (2021)}.

\bibitem{Eglinton22} J. Eglinton and K. Brandner, Geometric bounds on the power of adiabatic thermal machines, \href{https://doi.org/10.1103/PhysRevE.105.L052102}{Phys.\ Rev.\ E {\bf 105}, L052102 (2022)}.

\bibitem{Frim22prl} A.~G. Frim and M.~R. DeWeese, Geometric Bound on the Efficiency of Irreversible Thermodynamic Cycles, \href{https://doi.org/10.1103/PhysRevLett.128.230601}{Phys.\ Rev.\ Lett. {\bf 128}, 230601 (2022)}.

\bibitem{Sawchuk26} J.~R. Sawchuk and D.~A. Sivak, Global thermodynamic manifold for conservative control of stochastic systems, \href{https://doi.org/10.1103/j59j-q88v}{Phys.\ Rev.\ Research {\bf 8}, 013004 (2026)}.



\bibitem{Sivak12} D.~A. Sivak and G.~E. Crooks, Thermodynamic Metrics and Optimal Paths, \href{https://doi.org/10.1103/PhysRevLett.108.190602}{Phys.\ Rev.\ Lett. {\bf 108}, 190602 (2012)}. 


\bibitem{Brandner20} K. Brandner and K. Saito, Thermodynamic Geometry of Microscopic Heat Engines, \href{https://doi.org/10.1103/PhysRevLett.124.040602}{Phys.\ Rev.\ Lett. {\bf 124}, 040602 (2020)}.

\bibitem{Abiuso20} P. Abiuso and M. Perarnau-Llobet, Optimal Cycles for Low-Dissipation Heat Engines, \href{https://doi.org/10.1103/PhysRevLett.124.110606}{Phys.\ Rev.\ Lett. {\bf 124}, 110606 (2020)}.


\bibitem{Frim22} A.~G. Frim and M.~R. DeWeese, Optimal finite-time Brownian Carnot engine, \href{https://doi.org/10.1103/PhysRevE.105.L052103}{Phys.\ Rev.\ E {\bf 105}, L052103 (2022)}.

\bibitem{Li25} Z. Li and Y. Izumida, Decomposition of metric tensor in thermodynamic geometry in terms of relaxation timescales, \href{https://doi.org/10.1103/PhysRevE.111.034113}{Phys.\ Rev.\ E {\bf 111}, 034113 (2025)}.

\bibitem{Watanabe22} G. Watanabe and Y. Minami, Finite-time thermodynamics of fluctuations in microscopic heat engines, \href{https://doi.org/10.1103/PhysRevResearch.4.L012008}{Phys.\ Rev.\ Research {\bf 4}, L012008 (2022)}.


\bibitem{Miller19} H.~J.~D. Miller, M. Scandi, J. Anders, and M. Perarnau-Llobet, Work Fluctuations in Slow Processes: Quantum Signatures and Optimal Control, \href{https://doi.org/10.1103/PhysRevLett.123.230603}{Phys.\ Rev.\ Lett. {\bf 123}, 230603 (2019)}.

\bibitem{Miller20} H.~J.~D. Miller and M. Mehboudi, Geometry of Work Fluctuations versus Efficiency in Microscopic Thermal Machines, \href{https://doi.org/10.1103/PhysRevLett.125.260602}{Phys.\ Rev.\ Lett. {\bf 125}, 260602 (2020)}.



\bibitem{Risken_book} H. Risken, \href{https://doi.org/10.1007/978-3-642-61544-3}{{\it The Fokker--Planck Equation}, 2nd ed. (Springer, Berlin, 1988)}.



\bibitem{note:phitilde} The similarity-transformed eigenfunctions of the Hermitian operator are $\tilde{\phi}_i = e^{\Phi/2} \phi_i$, while the functions $e^{\Phi} \phi_i$ correspond to the left eigenfunctions of the Fokker-Planck operator since $L_{\mathrm{FP}}^\dagger = e^{\Phi}L_{\mathrm{FP}}e^{-\Phi}$.

\bibitem{note:Phi} The additive constant in $\Phi(x)$ has no effect on the spectrum of the Fokker-Planck operator. We keep it here to remain consistent with the conventional similarity transformation used in standard treatments (see, e.g., Ref.~\cite{Risken_book}).

\bibitem{note:sample} The larger scatter of the experimental data points in Fig.~\ref{fig:efficiency} at smaller $1/\tau$ may be due to the smaller number of samples. In the experiment of Ref.~\cite{Martinez16}, cycles were repeated for a fixed duration of $50$\,s at each value of $\tau$, so the number of samples decreases as $\tau$ increases.

\end{thebibliography}
\end{document}